\RequirePackage{lineno}
\documentclass[preprint,showpacs,preprintnumbers,amsmath,amssymb,superscriptaddress,aps,floatfix,prc]{revtex4-1}
\usepackage{longtable}
\usepackage{graphicx}% Include figure files
\usepackage{dcolumn}% Align table columns on decimal point
\usepackage{bm}% bold mathmwe
\usepackage[dvipsnames,usenames]{color}% This is just used temporarily, for notes to myself
\usepackage{subfigure}% Allow for grouped figures.
\usepackage{multirow}% Support for tables
\usepackage{amssymb}
\usepackage{mdwlist}% To use single-space itemize* and enumerate* lists
\usepackage{textcomp}

\def\today{\space\number\day\space\ifcase\month\or January\or February\or
    March\or April\or May\or June\or July\or August\or September\or October\or
    November\or December\fi\space\number\year}

\newcommand{\iso}[2]{$^{#1}$#2}

\begin{document}

\setpagewiselinenumbers
%\modulolinenumbers[5]
\linenumbers

\preprint{Draft 1.0}

\title{LUXSim: A Component-Centric Approach to Low-Background Simulations}

\author{D.~S. Akerib} %% Joined 4/2008
  \affiliation{Case Western Reserve University, Dept. of Physics, 10900 Euclid Ave, Cleveland OH 44106, USA}
\author{X. Bai} %% Joined 7/2009
  \affiliation{South Dakota School of Mines and technology, 501 East St Joseph St., Rapid City SD 57701, USA}
\author{S. Bedikian} %% Left 2/2010, Author until 2/2012
  \affiliation{Yale University, Dept. of Physics, 217 Prospect St., New Haven CT 06511, USA}
\author{E. Bernard} %% Joined January 2010
  \affiliation{Yale University, Dept. of Physics, 217 Prospect St., New Haven CT 06511, USA}
\author{A. Bernstein} %% From the start
  \affiliation{Lawrence Livermore National Laboratory, 7000 East Ave., Livermore CA 94551, USA}
%\author{A. Bolozdynya} %% From the start, left Case 7/09, Author until 7/11
%  \affiliation{National Research Nuclear University MEPHI, Faculty of the experimental and %theoretical physics, Kashirskoe sh.,31, Moscow, 115409, Russia}
\author{A. Bradley} %% From start
  \affiliation{Case Western Reserve University, Dept. of Physics, 10900 Euclid Ave, Cleveland OH 44106, USA}
\author{S.B. Cahn} %% Joined 2/09
  \affiliation{Yale University, Dept. of Physics, 217 Prospect St., New Haven CT 06511, USA}
\author{M.C. Carmona-Benitez} %% Joined 10/2009
  \affiliation{Case Western Reserve University, Dept. of Physics, 10900 Euclid Ave, Cleveland OH 44106, USA}
\author{D. Carr} %% From start
  \affiliation{Lawrence Livermore National Laboratory, 7000 East Ave., Livermore CA 94551, USA}
\author{J.J. Chapman} %% Joined 6/2007
  \affiliation{Brown University, Dept. of Physics, 182 Hope St., Providence RI 02912, USA}
\author{K. Clark} %% Joined 6/2008, Left 6/2010, author until 6/2012
  \affiliation{Case Western Reserve University, Dept. of Physics, 10900 Euclid Ave, Cleveland OH 44106, USA}
\author{T. Classen} %% Joined 10/2007
  \affiliation{University of California Davis, Dept. of Physics, One Shields Ave., Davis CA 95616, USA}
\author{T. Coffey} %% Joined 8/2010, Author on 8/2011
\affiliation{Case Western Reserve University, Dept. of Physics, 10900 Euclid Ave, Cleveland OH 44106}
%\author{A. Curioni} %% Left 6/2009, Author until 6/2011
%  \affiliation{Yale University, Dept. of Physics, 217 Prospect St., New Haven CT 06511, USA}
%\author{E. Dahl} %% From start; Left ???
%  \affiliation{Case Western Reserve University, Dept. of Physics, 10900 Euclid Ave, Cleveland OH 44106, USA}
\author{S. Dazeley} %% From start
  \affiliation{Lawrence Livermore National Laboratory, 7000 East Ave., Livermore CA 94551, USA}
%\author{A. Dobi} %% Joined 7/2011, Author on 7/2012
%  \affiliation{University of Maryland, Dept. of Physics, College Park MD 20742, USA}
%\author{B. Edwards} %% Joined 3/2011, Author on 3/2012.
%  \affiliation{Yale University, Dept. of Physics, 217 Prospect St., New Haven CT 06511, USA}
\author{L. de\,Viveiros} %% From the start 
  \affiliation{Brown University, Dept. of Physics, 182 Hope St., Providence RI 02912, USA}
\author{M. Dragowsky} %% Joined Jan 09
  \affiliation{Case Western Reserve University, Dept. of Physics, 10900 Euclid Ave, Cleveland OH 44106, USA}
\author{E. Druszkiewicz} %% From the start
  \affiliation{University of Rochester, Dept. of Physics and Astronomy, Rochester NY 14627, USA}
\author{C.H. Faham} %% Joined June 2007
  \affiliation{Brown University, Dept. of Physics, 182 Hope St., Providence RI 02912, USA}
\author{S. Fiorucci} %% From start
  \affiliation{Brown University, Dept. of Physics, 182 Hope St., Providence RI 02912, USA}
\author{R.J. Gaitskell} %% From start
  \affiliation{Brown University, Dept. of Physics, 182 Hope St., Providence RI 02912, USA}
\author{K.R. Gibson} %% Joined Oct 2009
  \affiliation{Case Western Reserve University, Dept. of Physics, 10900 Euclid Ave, Cleveland OH 44106, USA}
\author{C. Hall} %% Joined March 2008
  \affiliation{University of Maryland, Dept. of Physics, College Park MD 20742, USA}
\author{M. Hanhardt} %% Joined July 09
  \affiliation{South Dakota School of Mines and technology, 501 East St Joseph St., Rapid City SD 57701, USA}
\author{B. Holbrook} %% Joined December 2007
  \affiliation{University of California Davis, Dept. of Physics, One Shields Ave., Davis CA 95616, USA}
\author{M. Ihm} %% Joined December 2009
  \affiliation{University of California Berkeley, Department of Physics, Berkeley, CA 94720-7300, USA}
\author{R.G. Jacobsen} %% Joined December 2009
  \affiliation{University of California Berkeley, Department of Physics, Berkeley, CA 94720-7300, USA}
\author{L. Kastens} %% Joined June 2007
  \affiliation{Yale University, Dept. of Physics, 217 Prospect St., New Haven CT 06511, USA}
\author{K. Kazkaz\footnote{Corresponding author, kareem@llnl.gov}} %% From start
  \affiliation{Lawrence Livermore National Laboratory, 7000 East Ave., Livermore CA 94551, USA}
\author{R. Lander} %% Joined March 2008, Left 12/09, Author until 12/11.
  \affiliation{University of California Davis, Dept. of Physics, One Shields Ave., Davis CA 95616, USA}
\author{N. Larsen} %% Joined June 2010, Author on June 2011
  \affiliation{Yale University, Dept. of Physics, 217 Prospect St., New Haven CT 06511, USA}
\author{C. Lee} %% Joined August 2009
  \affiliation{Case Western Reserve University, Dept. of Physics, 10900 Euclid Ave, Cleveland OH 44106, USA}
\author{D. Leonard} %% Joined November 2008, Left February 2010, Author until February 2012.
  \affiliation{University of Maryland, Dept. of Physics, College Park MD 20742, USA}
\author{K. Lesko} %% Joined March 2007
  \affiliation{Lawrence Berkeley National Laboratory, 1 Cyclotron Rd., Berkeley CA 94720, USA}
%\author{A. Lindote} %% Joined December 2010, Author on December 2011.
%  \affiliation{LIP-Coimbra, Department of Physics, University of Coimbra, Rua Larga, 3004-516 Coimbra, Portugal}
%\author{M.I. Lopes} %% Joined December 2010, Author on December 2011.
%  \affiliation{LIP-Coimbra, Department of Physics, University of Coimbra, Rua Larga, 3004-516 Coimbra, Portugal}
\author{A. Lyashenko} %% Joined July 2009
  \affiliation{Yale University, Dept. of Physics, 217 Prospect St., New Haven CT 06511, USA}
\author{D.C. Malling} %% Joined August 2007 
  \affiliation{Brown University, Dept. of Physics, 182 Hope St., Providence RI 02912, USA}
\author{R. Mannino} %% Joined June 2009
  \affiliation{Texas A \& M University, Dept. of Physics, College Station TX 77843, USA}
\author{D.N. McKinsey} %% Joined June 2007
  \affiliation{Yale University, Dept. of Physics, 217 Prospect St., New Haven CT 06511, USA}
\author{D.-M Mei} %% Joined July 2008
  \affiliation{University of South Dakota, Dept. of Physics, 414E Clark St., Vermillion SD 57069, USA}
\author{J. Mock} %% Joined August 2007
  \affiliation{University of California Davis, Dept. of Physics, One Shields Ave., Davis CA 95616, USA}
\author{M. Morii} %% Joined June 09, Left May 2011, Author until May 2013
  \affiliation{Harvard University, Dept. of Physics, 17 Oxford St., Cambridge MA 02138, USA}
\author{H. Nelson} %% Joined November 2008
  \affiliation{University of California Santa Barbara, Dept. of Physics, Santa Barbara, CA, USA}
%\author{F. Neves} %% Joined December 2010, Author on December 2011.
%  \affiliation{LIP-Coimbra, Department of Physics, University of Coimbra, Rua Larga, 3004-516 Coimbra, Portugal}
\author{J.A. Nikkel} %% Joined April 2008
  \affiliation{Yale University, Dept. of Physics, 217 Prospect St., New Haven CT 06511, USA}
\author{M. Pangilinan} %% Joined Feb 2010
  \affiliation{Brown University, Dept. of Physics, 182 Hope St., Providence RI 02912, USA}
\author{P.D. Parker} %% Joined 9/2010, Author on 9/2011
  \affiliation{Yale University, Dept. of Physics, 217 Prospect St., New Haven CT 06511, USA}
\author{P. Phelps} %% Joined June 2008
  \affiliation{Case Western Reserve University, Dept. of Physics, 10900 Euclid Ave, Cleveland OH 44106, USA}
%\author{J. Pinto da Cunha} %% Joined December 2010, Author on December 2011.
%  \affiliation{LIP-Coimbra, Department of Physics, University of Coimbra, Rua Larga, 3004-516 Coimbra, Portugal}
\author{T. Shutt} %% From start
  \affiliation{Case Western Reserve University, Dept. of Physics, 10900 Euclid Ave, Cleveland OH 44106, USA}
%\author{C. Silva} %% Joined December 2010, Author on December 2011.
%  \affiliation{LIP-Coimbra, Department of Physics, University of Coimbra, Rua Larga, 3004-516 Coimbra, Portugal}
\author{W. Skulski} %% From start
  \affiliation{University of Rochester, Dept. of Physics and Astronomy, Rochester NY 14627, USA}
%\author{V.N. Solovov} %% Joined December 2010, Author on December 2011.
%  \affiliation{LIP-Coimbra, Department of Physics, University of Coimbra, Rua Larga, 3004-516 Coimbra, Portugal}
\author{P. Sorensen} %% From start
  \affiliation{Lawrence Livermore National Laboratory, 7000 East Ave., Livermore CA 94551, USA}
\author{J. Spaans} %% Joined February 2009
  \affiliation{University of South Dakota, Dept. of Physics, 414E Clark St., Vermillion SD 57069, USA}
\author{T. Stiegler} %% Joined October 2007
  \affiliation{Texas A \& M University, Dept. of Physics, College Station TX 77843, USA}
\author{R. Svoboda} %% From start
  \affiliation{University of California Davis, Dept. of Physics, One Shields Ave., Davis CA 95616, USA}
\author{M. Sweany} %% From start, Left 6/11, Author until 6/13
  \affiliation{University of California Davis, Dept. of Physics, One Shields Ave., Davis CA 95616, USA}
\author{M. Szydagis} %% Joined May 2010, Author on May 2011
  \affiliation{University of California Davis, Dept. of Physics, One Shields Ave., Davis CA 95616, USA}
\author{J. Thomson} %% Joined May 2008
  \affiliation{University of California Davis, Dept. of Physics, One Shields Ave., Davis CA 95616, USA}
\author{M. Tripathi} %% From start
  \affiliation{University of California Davis, Dept. of Physics, One Shields Ave., Davis CA 95616, USA}
\author{J.R. Verbus} %% Joined July 09
  \affiliation{Brown University, Dept. of Physics, 182 Hope St., Providence RI 02912, USA}
\author{N. Walsh} %% Joined February 2008
  \affiliation{University of California Davis, Dept. of Physics, One Shields Ave., Davis CA 95616, USA}
\author{R. Webb} %% Joined April 2008
  \affiliation{Texas A \& M University, Dept. of Physics, College Station TX 77843, USA}
\author{J.T. White} %% From start
  \affiliation{Texas A \& M University, Dept. of Physics, College Station TX 77843, USA}
\author{M. Wlasenko} %% Joined October 2009, Left May 2011, Author until May 2013
  \affiliation{Harvard University, Dept. of Physics, 17 Oxford St., Cambridge MA 02138, USA}
\author{F.L.H. Wolfs} %% From start
  \affiliation{University of Rochester, Dept. of Physics and Astronomy, Rochester NY 14627, USA}
\author{M. Woods} %% Joined October 2008
  \affiliation{University of California Davis, Dept. of Physics, One Shields Ave., Davis CA 95616, USA}
\author{C. Zhang} %% Joined February 2009
  \affiliation{University of South Dakota, Dept. of Physics, 414E Clark St., Vermillion SD 57069, USA}
     
\date{\today}% It is always \today, today,

\begin{abstract}
Geant4 has been used throughout the nuclear and high-energy physics community to simulate energy depositions in various detectors and materials. These simulations have mostly been run with a source beam outside the detector. In the case of low-background physics, however, a primary concern is the effect on the detector from radioactivity inherent in the detector parts themselves. From this standpoint, there is no single source or beam, but rather a collection of sources with potentially complicated spatial extent. LUXSim is a simulation framework used by the LUX collaboration that takes a component-centric approach to event generation and recording. A new set of classes allows for multiple radioactive sources to be set within any number of components at run time, with the entire collection of sources handled within a single simulation run. Various levels of information can also be recorded from the individual components, with these record levels also being set at runtime. This flexibility in both source generation and information recording is possible without the need to recompile, reducing the complexity of code management and the proliferation of versions. Within the code itself, casting geometry objects within this new set of classes rather than as the default Geant4 classes automatically extends this flexibility to every individual component. No additional work is required on the part of the developer, reducing development time and increasing confidence in the results. We describe the guiding principles behind LUXSim, detail some of its unique classes and methods, and give examples of usage.
\end{abstract}

% 32.50.%61.25.Bi Liquid noble gases
\maketitle

%
%	Introduction
%
\section{Introduction}
\label{s:Intro}

Geant4 is a physics process simulation package developed at CERN, initially for high-energy physics simulations~\cite{Agostinelli2003,Allison2006,Geant4}. In the majority of high-energy experiments, the primary particles are generated separate from the active detector elements. This provided a clean distinction in the simulation between the machinery used to generate the beam and the hardware used to measure the beam's effects.

Over the years, Geant4 has been expanded to make it more useful for experiments at nuclear energies, including the category of low-background experiments such as neutrino research, searches for neutrinoless double-beta decay, and searches for WIMP Dark Matter. This expanded functionality included additional code to handle electromagnetic interactions down to 250 eV in energy, neutron interactions down to thermal energies, radioactive decays, and event generation from an arbitrary volume rather than a point or a beam.

Historically, a Geant4 simulation of a low-background experiment would be run, recording energy depositions only from the active detector components. Inevitably, unexpected phenomena required recording data from passive components as well, to account for all energy released in an event. Regardless of the time of an interaction, additional code had to be written into the simulation for every component that recorded data. It was rarely known {\it a priori} which parts were altering the observed energy depositions, so more and more components had to be included in the data record, leading to a large proliferation of additional code within the simulation.

In addition to data recording, low-background experiments must pay special attention to the energy sources of each individual component and material within the detector, support structure, shielding, and environment. These sources include cosmic ray spallation, intrinsic radioactivity, and surface contaminants, and multiple sources are frequently required for a single component. Although educated guesses could be made, it is difficult to know beforehand which sources in which components are the most relevant to the experiment. Sources therefore have to be added to more and more components, with additional code required for each combination.

In the end, it is much easier to simply ensure that {\it all} parts have the ability to record data and carry multiple radioactive loads. The code to handle data recording and the code to handle intrinsic radioactivity is largely independent of the part itself. This implies the need for a set of classes that provide a consistent approach to both requirements. This paper includes details on such a new set of classes.

The new features described in this paper is useful across multiple current and future experiments involving nuclear-scale energies and low levels of background activity. They were therefore developed into a generalized code base called LUXSim. These features include creating multiple, simultaneous primary particle types and composite sources, as well as allowing those particles to be generated from multiple volumes of arbitrary spatial extent. In addition to these physics-motivated features, LUXSim has a set of guiding principles to increase reliability and reproducibility, and to reduce the time and effort required to use or expand on the package.

In Section~\ref{s:LUX}, we briefly cover the LUX experiment to provide context for the simulation package, and in Section~\ref{s:GuidingPrinciples} we describe the guiding principles for LUXSim. In Section~\ref{s:Subsystems} we discuss the details of the subsystems that make up the Geant4 user code within LUXSim. In Section~\ref{s:DataProcessing} we describe how the resulting data can be post-processed to make it similar to the data stream coming from the physical electronics. In Section~\ref{s:Exercise} we exercise the basic functionality of LUXSim, comparing simulation data with experimental data from a single-phase detector, as well as making preliminary predictions relevant for optical photon collection. In Section~\ref{s:Expansion} we describe how to use the LUXSim infrastructure for other experiments.

%
%	LUX
%
\section{The LUX Experiment}
\label{s:LUX}

LUX is a search for WIMP Dark Matter based at the Sanford laboratory in Lead, South Dakota~\cite{McKinsey2010,LUX2011}. LUX utilizes a dual-phase detector with a 300-kg liquid xenon target (100 kg fiducial mass) to obtain a projected sensitivity in the WIMP-nucleon elastic scattering cross section of $7 \times 10^{-46}$ cm$^2$ for a 100-GeV WIMP. To attain this level of sensitivity, there can only be 2 background events in the 5-25~keV region of interest after 300~days of running, qualifying LUX as a low-background experiment.

\begin{figure}[t]
\centering
\includegraphics[width=5in]{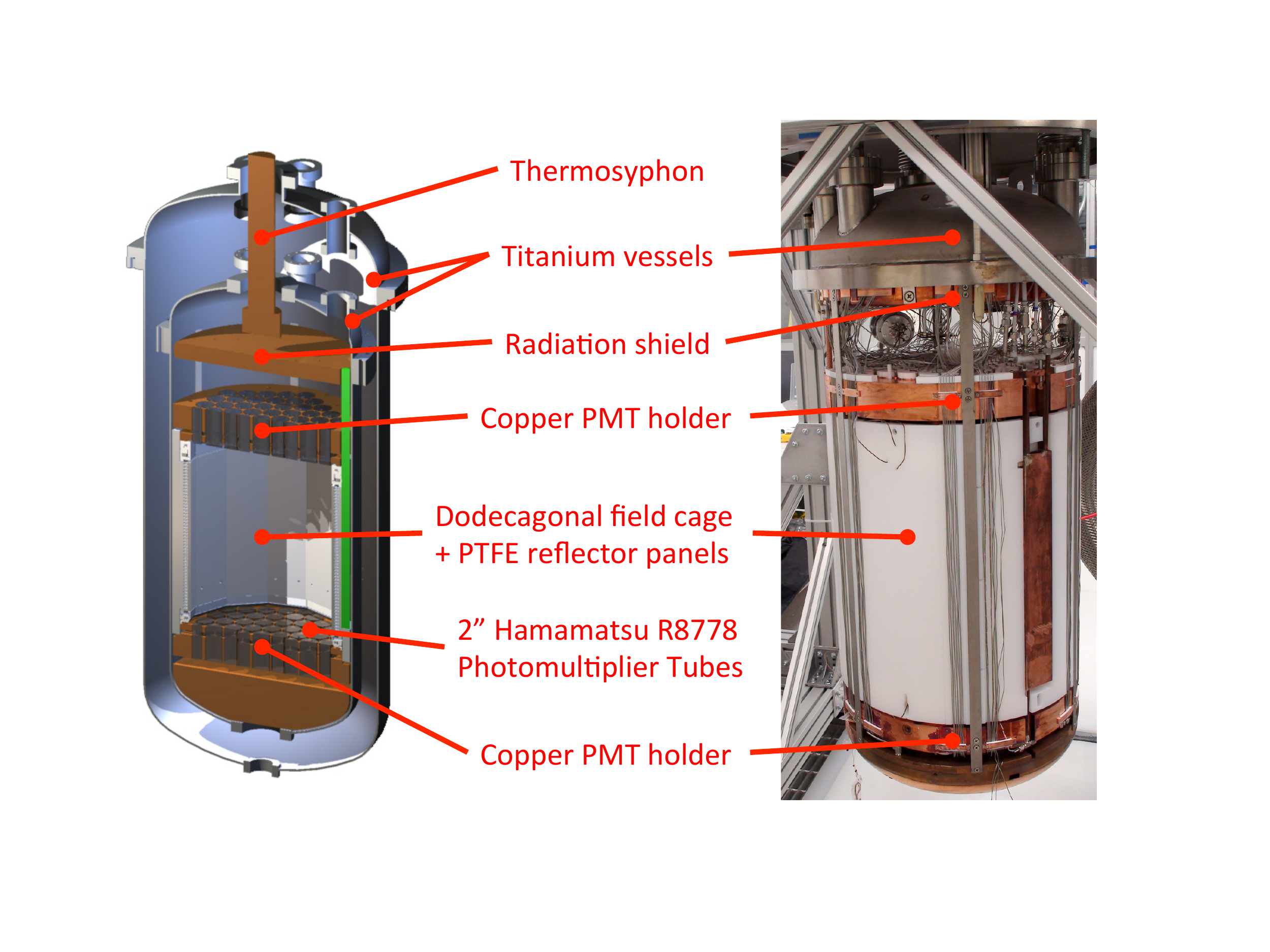}
\parbox{5in}{\vspace{5pt}\caption{\small{Photo of the LUX detector, and an engineering rendering. Only the top of the outer titanium vessel is present in the photo.}}
\label{fig:LUXDetector}}
\end{figure}

The detector is comprised of a titanium cryostat inside a titanium vacuum vessel. The photomultiplier tubes used to detect the scintillation light resulting from charged particle interactions are housed in monolithic copper frames. The LUX detector will be installed in an 8-meter-diameter water tank to provide shielding from external gammas and neutrons. This water tank will be instrumented with photomultiplier tubes to create a  tag for muons that pass close to the active xenon volume. The water tank also thermalizes and captures neutrons, and by adding gadolinium to the water, the neutron-tagging efficiency is increased because of the resulting 8-MeV gamma cascade. Figure~\ref{fig:LUXDetector} shows the LUX detector itself.

A particle interaction in the LUX detector generates two signals. The first is the primary flash of scintillation light created during the initial xenon or electron recoil, referred to as the S1 light. The recoil also creates ionization, and those liberated electrons are drifted up by an electric field to a gaseous volume just below the top bank of photomultiplier tubes. A field gradient across the liquid / gas boundary extracts the drifted electrons from the liquid surface, and a high electric field within the gaseous volume causes the drifting electrons to create scintillation light, producing a second scintillation pulse referred to as S2. A detailed analysis of this sort of signal from dual-phase detectors has been studied by previous WIMP searches such as {\sc Xenon10}~\cite{Aprile2011,Angle2008} and {\sc Zeplin-III}~\cite{Akimov2006,Lebedenko2009}.

%
%	Guiding principles
%
\clearpage
\section{The LUXSim Guiding Principles}
\label{s:GuidingPrinciples}

Basic physics features of LUXSim were listed in Section~\ref{s:Intro}, but there are other features that should be incorporated to make the Monte Carlo simulation code itself easier to use and develop. In this section, we cover the desired feature set for LUXSim and how these features are implemented.

\subsubsection{Keep LUXSim simple for users}

LUXSim is developed primarily by a subsection of the LUX collaboration, but with the intent that anyone in the collaboration can use the simulation to produce results on a very short time scale. LUXSim is therefore controlled mainly through the use of macro commands issued at run time, rather than recoding and recompiling. Because lower-level coding and recompilations are kept to a minimum, there is less chance for an error to make its way into the code base.

LUXSim, like the underlying Geant4 framework, is not intended to handle all conceivable situations ``out of the box''. Users are encouraged, however, to request new features. If those features are simple to build into the code and do not interfere with the performance and results of the current code base, they are implemented. If those features require a more fundamental code change, they are evaluated for universality---if an appreciable number of users would make use of the new feature, it is implemented. Ultimately, to handle very specific cases that are not a standard part of LUXSim, users work with the developers to ensure quality results.

A directory hierarchy was created to allow for conceptual segregation of the LUXSim subsystems. In the examples of user code distributed with the Geant4 package, all source files are contained in a single directory called ``src'', and all header files in an ``include'' directory. Within LUXSim, there are approximately four dozen code files, and placing them all in a single directory would make conceptual separation difficult. LUXSim files are therefore spread across separate ``generator'', ``geometry'', ``io'', ``management'', ``physicslist'', and ``processing'' directories. Each directory contains its own ``src'' and ``include'' directories.

\subsubsection{Scalability}

Because it is based on Geant4, LUXSim is a C++ program, allowing for highly object-oriented programming techniques. LUXSim relies heavily on subclassing and multiple instances. For example, LUXSim has a generic detector class from which all possible geometries inherit. This allows for one geometry to be quickly and easily swapped in for another at run time.

We also made heavy use of C++ container classes, to allow for scalability to meet current and future needs within the simulation. To this end, we employ vectors, strings, and stringstreams wherever possible, and have restricted the use of hard-coded array sizes and character arrays with a finite size.

\subsubsection{Reproducibility}

From time to time, two ostensibly similar Geant4 simulations can produce very different results. These differences can be the result of various discrepancies between the simulations, e.g, in the geometry, the material properties, the particle interaction models, or the generation mechanism for primary particles. It can be extremely difficult to reproduce those simulations to identify the underlying cause of the differences, especially if months or years have passed since the programs were run.

We have therefore built automatic record-keeping into the LUXSim output file. Each file contains a text header that contains valuable pieces of information required to know exactly how the data in the file was generated. That information includes which computer ran the simulation, its operating system, the version of Geant4 used, a time/date stamp, and the seed used to initialize the random number generator. Within the LUX collaboration, we use the Subversion (``SVN'') software management system~\cite{Subversion2010}, and the output file header includes not only the specific LUXSim SVN version, but any differences between the code that ran the simulation and the code checked in under that version in the software repository (i.e., the SVN ``diffs'').

The header information is written to every output file to eliminate concerns over keeping separate log files associated with the individual simulation data files. The header size can depend greatly on the amount of code that may have been changed between the files in the central SVN repository and the files on the local hard drive, but rarely exceeds more than a few kilobytes.

\subsubsection{Agnostic binary output format}

The data generated with LUXSim are recorded to a binary file with a custom, well-documented format, described in Section~\ref{ss:DataRecording}. LUXSim users can employ any software package to analyze the simulation data. LUXSim includes routines to read the data files using either ROOT~\cite{Brun1996} or Matlab~\cite{Matlab2010}. This allows members of the LUX collaboration to use either package for analysis and processing. Neither package is required for the simulation to run since LUXSim does not use any ROOT or Matlab libraries or classes within the simulation code.

\subsubsection{Self-registering objects}

Within LUXSim, there is a manager class called LUXSimManager. This class contains registration methods that all classes within LUXSim use. The manager contains a list of all pointers to all subsystems within LUXSim, and handles the communications between the various parts of the simulation. As such, pointers do not have to be explicitly passed from object to object, and all parts of the simulation have automatic access to information from any class within LUXSim via the LUXSimManager.

The classes within LUXSim must therefore respect the automatic registration with LUXSimManager. This automatic registration is part of the built-in framework, and simply inheriting from various LUXSim classes performs the job without requiring additional code. Additional details of the manager are discussed in Section~\ref{ss:Manager}.

\subsubsection{Component-centric approach to event generation and recording}

LUXSim utilizes a custom class called LUXSimDetectorComponent. This class itself inherits from the Geant4 class G4PVPlacement, and is therefore intimately associated with the geometry of any given detector. By recasting all physical volumes as LUXSimDetectorComponents, we ensure that all components automatically have access to the LUXSim infrastructure.

Part of this infrastructure includes methods for each component to store its own radioactive identities and rates. The components can also store their own record levels, so that specific information about particles passing through or interacting within the volume can be recorded, or not, as specified by the user at run time. While the LUXSim code base cannot possibly anticipate every question a user may want to answer by running a simulation, we have attempted to build in a great deal of flexibility so that users can decide for themselves whether they are interested in energy depositions either within the active detector components or some nearby, inert, supporting structure.

In Geant4, an ``event'' is all the interactions deriving from the full complement of primary particles that are generated in a single loop. The primary particles can themselves be radioactive nuclei, which have finite lifetimes. Because of the stochastic nature of particle decays, it is entirely possible for interactions from one event to occur earlier in time that interactions from a previous event (see Section~\ref{sss:UThDecayChains}).

Details of the LUXSimDetectorComponent class are available in Sections~\ref{ss:Geometry}, \ref{ss:Generator}, and \ref{ss:DataRecording}.

%
%	LUXSim Subsystems
%
\section{LUXSim Subsystems}
\label{s:Subsystems}

LUXSim uses a component-centric approach to creating a Geant4-based simulation, which means the detector components themselves store their own levels of radioactivity and step-by-step energy depositions. The internal management and information flow, however, cannot be tied to the components because they change from geometry to geometry. LUXSim therefore makes use of a manager class to handle inter-class communications.

\begin{figure}[b]
\centering
\includegraphics[width=15cm]{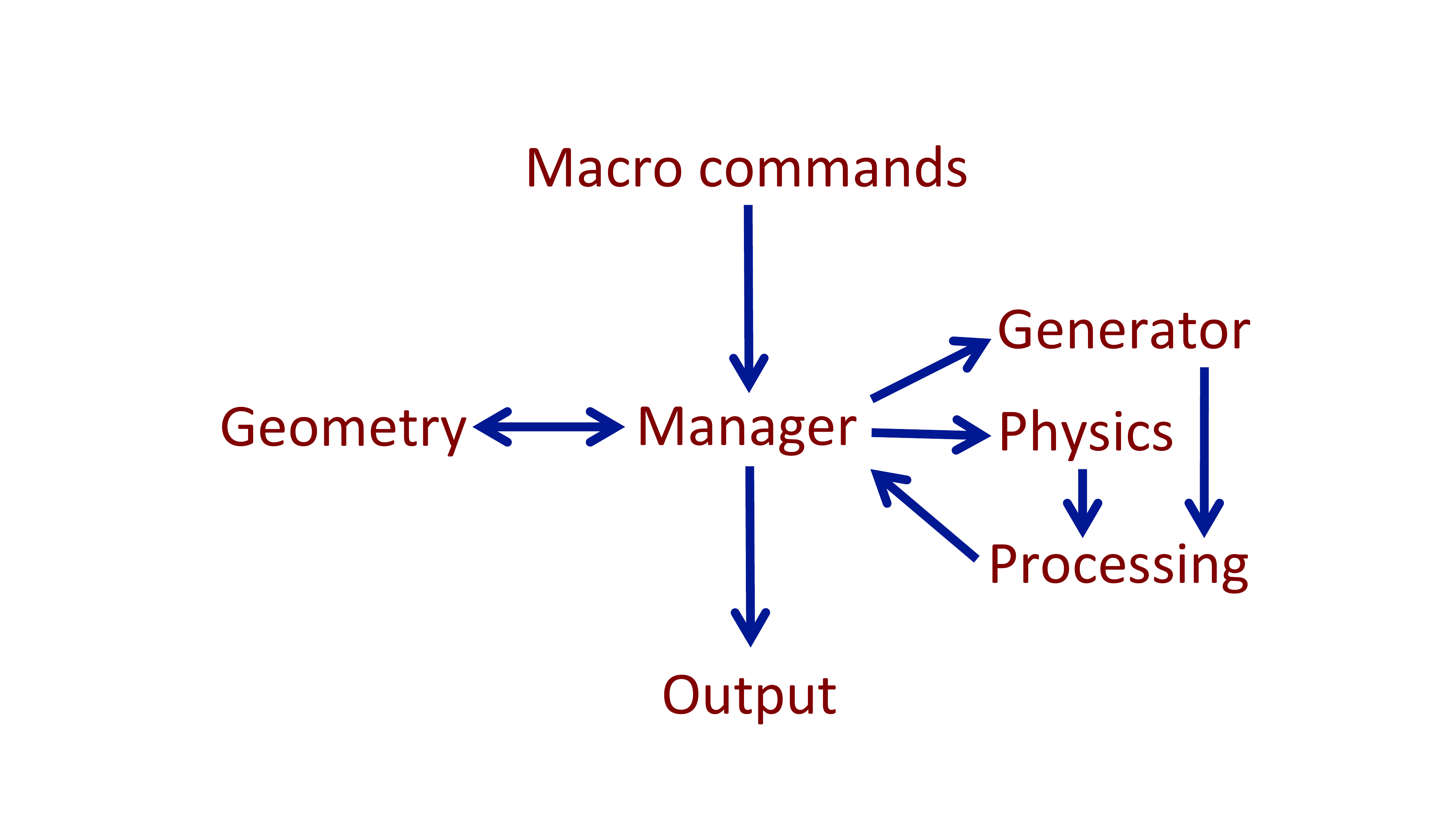}
\parbox{5in}{\vspace{5pt}\caption{\small{Information flow in LUXSim. The manager handles all communication between subsystems, while the detector components in the geometry subsystem hold the vital information.}}
\label{fig:InfoFlow}}
\end{figure}

The flow of information is shown in Fig.~\ref{fig:InfoFlow}. The user controls the simulation via macro commands, which are processed by the manager. Before the simulation actually begins, the manager performs final calculations and bookkeeping based on the latest user commands.

In this section, we detail the subsystems within LUXSim.

%%%%%
\subsection{The LUXSim Manager Class}
\label{ss:Manager}

The LUXSimManager class sits in the middle of the information flow within LUXSim. It contains registration methods for all the other LUXSim classes, and contains access methods so that any subclass can retrieve those pointers. In turn, all the other classes store a pointer to the manager singleton to provide two-way communication within the simulation. These registration methods are in the class constructors so that any objects that inherit from these LUXSim classes automatically become a part of the larger framework.

The manager class processes user commands, sets sources and record levels within the detector components, sets flags and parameters for the physics models, and provides control over the randomization seed. Once the simulation parameters are set, it performs final calculations regarding source strength ratios, creates and makes accessible the reproducibility information stored in the header, and assigns integer indices to the volume names to reduce the size of the output file.

At the beginning of the simulation, the manager appends a ``.tmp'' extension to the output file name. In addition, it keeps a vector list of all detector components, as well as the sum total of all radioactivity within each component. The manager also records the primary particle information for every event, which is very important if the particle is randomly selected from an extended decay chain or generated from anywhere within a three-dimensional object. The manager controls which physical volume will contain the next event based on the pre-calculated ratios of activity. While the simulation is running, step information is passed along to the manager for it to parcel out to the correct detector component object. Finally, if the simulation ends cleanly, the manager removes the ``.tmp'' extension to signify a complete and whole data file.

If the simulation does not end cleanly, the user may use routines specifically written to recover as many full, reliable events as may have been recorded in the data file. In this case, however, the number of events contained in the data file would have to be adjusted, and the header file of the cleaned data set re-written. Because of the uncertainty surrounding incomplete datafiles, users are encouraged to simply re-run the simulation.

%%%%%
\subsection{Geometry and LUXSim Detector Components}
\label{ss:Geometry}

Rather than using the Geant4 class G4PVPlacement for instantiating physical volumes, every physical volume in LUXSim has been recast as a LUXSimDetectorComponent, which inherits from G4PVPlacement. This new class has the features that satisfy the requirements for consistent treatment of data recording and event generation described in Section~\ref{s:Intro}.

The LUXSimDetectorComponent class contains the record level associated with any particular physical volume. The record level determines the amount of information recorded to the output file (see Section~\ref{ss:DataRecording}). The user can specify the record level for any component with a simple macro command, thus choosing at run time which components will serve as particle ``detectors'' rather than hard-coding Geant4-style sensitive detectors. The record levels are covered in detail in Section~\ref{ss:DataRecording}.

A LUXSimDetectorComponent can also have an arbitrary number of radioactive sources associated with it, each with an individual activity level. At the beginning of the run, the /LUXSim/beamOn macro command calculates the total activity in all components. During the running of the simulation, the manager randomly chooses both the component and the source within the component, all weighted by the specified activities. The timing, energy, momentum, and secondary particle chains for each event are handled by a variety of generators, described in Section \ref{ss:Generator}. Additionally, a volume-sampling method is used to determine the starting position for the primary particles. The detector component's center coordinates in the global reference frame are calculated, as well as its spatial extent in the x, y, and z directions and its orientation within the global reference frame. The vertex location of the event is passed to the PrimaryGeneratorAction, which then passes the simulation processing to the internal Geant4 libraries.

This type of component-centric approach to event generation and information recording makes heavy use of the two-way communications between the geometry subsystem and the LUXSimManager class. The macro commands for geometry construction, radioactive and particles sources, and detector record levels are passed to the manager. Before the simulation starts, those settings are accessed by the geometry sub-system, and stored in the LUXSimDetectorComponent classes. That stored information is then accessed through the manager by other sub-systems: source choices are passed to the PrimaryGeneratorAction, record level choices are passed to the input/output subsystem, and so on.

Four default detector geometries are available with various options, including LUX1.0 (the full detector system), LUX0.1 (a prototype detector for studying engineering, xenon liquification and purification, and signal handling), the LLNL single-phase detector (see Section~\ref{ss:LLNLSinglePhase}), and an empty LUX cryostat to study the effects of the water shield without the complex internal structure of the LUX1.0 detector. The LUX1.0 geometry was designed with two purposes: providing a background model from both internal detector components and external sources, and determining the light response of the detector. The simulated LUX1.0 geometry is based on all available engineering diagrams and specifications, ensuring the highest possible confidence in the simulation results.

\begin{figure}[b]
\centering
\includegraphics[width=5cm]{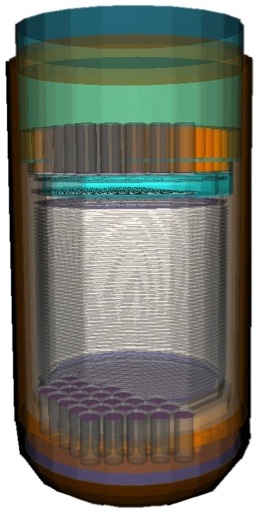}
\parbox{5in}{\vspace{5pt}\caption{\small{The LUXSim rendering of the internal components of the LUX1.0 detector. The simulated detector includes the major components that affect the optical response and background, including the Hamamatsu R8778 PMTs, the reflective PTFE surfaces, and the grid wires and frames. All 122 PMTs are present in the simulation, although the visualization hides some of them. Compare to Fig.~\ref{fig:LUXDetector}.}}
\label{fig:LUXSim1_0Det}}
\end{figure}

\begin{figure}[t]
\centering
\includegraphics[width=10cm]{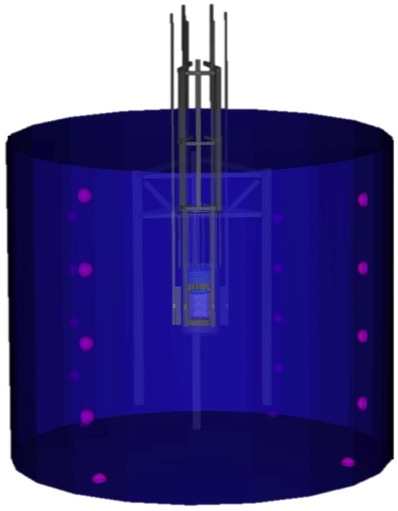}
\parbox{5in}{\vspace{5pt}\caption{\small{LUXSim rendering of the LUX1.0 detector inside the water shield. The water shield has 20 ten-inch Hamamatsu R7081 PMTs for the Muon Veto System. Also visible in this image are the guide tubes for the calibration source, located just outside of the cryostat.}}
\label{fig:LUXSimWaterDet}}
\end{figure}

To be able to select the different detectors at run time with macro commands, each detector inherits from a base LUXSimDetector class. The outermost volume of the specific detector (e.g, the cryostat of the LUX1.0 detector, or the vacuum vessel of the LLNL single-phase detector) is incorporated in the simulation as a LUXSimDetectorComponent after the user choice is made, but before the simulation starts running. These geometries can exist as stand-alone detectors, or encased inside the LUX water tank. See Figs.~\ref{fig:LUXSim1_0Det} and \ref{fig:LUXSimWaterDet} for images of the LUX1.0 detector.

Dual-phase detectors incorporate wire grids and meshes to define the electric field volumes. Within the LUX1.0 geometry, these thousands of wires are placed individually to provide accurate handling of the optical photons. There is, however, an option to remove grid wires from the LUX1.0 and LUX0.1 detectors for simulations that do not require optical physics.

Whenever multiple copies of similar components exist, component classes are used to avoid code duplication. One example is  the grid class used for both the LUX1.0 and LUX0.1 detectors. A single call to this grid class, utilizing appropriate dimensions and parent volumes as parameters, can create grids for multiple detectors. A second example is the photomultiplier tubes, where a single tube geometry is created, and multiple instances created as necessary to fully populate the detector. This approach follows established Geant4 practices.

%%%%%
\clearpage
\subsection{Optical properties}
\label{ss:OpticalProperties}

Energy deposition in the LUX detector will be measured in both scintillation and ionization channels. Both of these channels are measured using scintillation light. The scintillation channel is measured with primary scintillation (S1), and the ionization channel is measured with secondary scintillation (S2). The S1 and S2 light collection efficiency must therefore be very well understood to properly calibrate the detector. The collection efficiency value is not constant, but is a function of effects such as position, material reflectivity, and scattering and absorption lengths in the xenon.

Because of the low-energy deposition that characterizes a possible WIMP signal, the amount of scintillation light generated may be quite small. LUXSim must therefore handle scintillation light collection down to the single-photon level, making optical modeling of paramount importance. The LUX experiment needs to detect the S1 and S2 signals from recoils within the liquid xenon target, and LUXSim needs to be able to accurately reproduce the spatial and timing characteristics of those pulses. Twin arrays of Hamamatsu R8778 PMTs detect the scintillation light, which is narrowly peaked at 178~nm. Light collection is enhanced by fully encompassing the active region with sheets of PTFE, which has a high reflectivity for UV photons. Detailed modeling of the effects of these materials is crucial in LUXSim.

The optical properties of liquid xenon at 178~nm have been studied in several different experiments \cite{Baldini2006}. One critical parameter of liquid xenon is the refractive index, which the {\sc Xenon10} collaboration set at 1.69 at 178~nm. This value was combined with data on the liquid xenon refractive index from 361.2 to 643.9~nm~\cite{Sinnock1969}, and a polynomial fit used to interpolate between the points. Another critical optical parameter is the Rayleigh (lossless) scattering length, which has been measured to be 30~cm \cite{Ishida1997,Seidel2002}.

A third critical optical parameter of liquid xenon is the absorption length of its own scintillation light. Pure liquid xenon is expected to be transparent to its own scintillation light \cite{Baldini2005}. Loss mechanisms are dominated by impurities, including in particular oxygen and water \cite{CfAAbsorption}. The expected absorption coefficients for these impurities are convolved with the xenon scintillation spectrum, and the characteristic absorption length is fit to the result for a given impurity concentration. The LUX getter is rated for impurity removal below the ppb level; this has led to a conservative estimate of a minimum absorption length of 100~m, which is used as the LUXSim default.

PTFE reflectivity is very important for estimates of overall light collection in LUX, as well as light collection dependence on event position. Because any given photon can reflect many times off the PTFE walls, a relatively small change in the PTFE reflectivity can result in a large change in the final light collection efficiency. Measurements performed on PTFE samples from a variety of preparation and treatment types show a very large variation in their overall reflectivities, as well as variations in their specular and diffuse responses \cite{Silva2010a,Silva2010b}. LUX PTFE samples were measured in gas to determine their predicted reflectivity, with the results implemented in LUXSim. Separate reflectivity values are implemented at the liquid and gas xenon interfaces, as PTFE reflectivity in liquid xenon is predicted to increase substantially from that measured in gas due to the change in refractive index of the incident medium. The reflectivity model used in LUXSim for PTFE is 100\% diffuse, although measurements from Ref.~\cite{Silva2010a} indicate that the specular component of LUX PTFE increases with increasing angle of incidence. Note that in this paper, the angle of incidence is measured with respect to normal. The effects of diffuse and specular reflection models and the default values used for PTFE reflectivity in LUXSim are discussed in Section~\ref{ss:LightCollection}.

The geometry includes the electric field grids used within the active region. The LUX field grids are strung steel wire, with 98-99\% geometric transparency at $0^\circ$ incidence with respect to normal. The exception is the anode grid, which is a mesh of 88\% transparency at normal incidence. The grids are modeled as individual wires, as their geometric transparency decreases with increasing incidence angle. Very little data exists for the reflectivity of steel at 178~nm in a high refractive index medium. LUXSim uses a conservative 10\% reflectivity for baseline estimates of light collection efficiency; studies of LUX light collection using LUXSim with default optical parameters have shown that a high grid reflectivity of 100\% can boost light collection by $\sim$10\% for any given PTFE reflectivity.

The LUX R8778 PMTs feature a fused silica window in front of the photocathode. This feature is implemented in LUXSim in order to include the effects of photon reflection from the quartz windows, which is particularly important for photons incident on the PMT windows at high angles of incidence. Values of refractive index at UV wavelengths are found in \cite{Malitson1965}, with temperature-dependent changes calculated using \cite{Matsuoka1991}. Results yield an expected refractive index of 1.59 at 178~nm.

LUXSim also includes optical properties defined at lower wavelengths appropriate for water Cerenkov processes. These properties facilitate the study of the optical response of the muon veto implemented in the LUX water shield. Refractive indices for water are taken from \cite{Huibers1997}; absorption lengths are taken from several sources \cite{Quickenden1980,Sogandares1997,Pope1997}.

%%%%%
\subsection{The event generator subsystem}
\label{ss:Generator}

The LUXSimSource class provides a general framework for all the event generators in LUXSim. Currently, the list of generators includes all single-radionuclide decays, \iso{238}{U} and \iso{232}{Th} radioactive decay chains, an AmBe source with neutrons and associated gammas, \iso{252}{Cf} fission neutrons and gammas, and a cosmic muon generator with spallation neutrons. Each source inherits from the base class and sets its own values for the particle type, energy, and direction. A vector of all the source types is kept in a separate class, the LUXSimSourceCatalog, and is used for setting sources in materials and generating events. The LUXSimPrimaryGeneratorAction class can handle both LUXSim-style sources specified through macro commands, or generate an event defined by standard Geant4 General Particle Source (G4GPS) commands. The user can remove all sources and redefine their activity levels without requiring a recompilation or restart of LUXSim, which allows for serial processing of multiple simulations without quitting the program. This feature is useful on computer clusters with a managed job queueing, where it may take some time for the job to start.

To add a source to the simulation, the user specifies the component name, and the name and activity of the generating source to put in the component. For example, the command to load a 100~mBq/kg AmBe source on a specific PMT window is ``\texttt{/LUXSim/source/set Bottom\_PMT\_Window\_39 AmBe 100 mBq/kg}''. The volume names are keyed to the physical volume name as set in the LUXSimDetectorComponent declaration. Activity units of Ci or Bq with standard SI prefixes are accepted, while the per unit mass is optional. For the case of the activity being defined per unit mass, the mass itself is automatically calculated based on the Geant4 geometry and material density. The age of the source can also be specified at the end of the command (e.g. for the ``\texttt{U238Chain}'' generator, one might add ``\texttt{2~Gyr}''). Radioactive loads can be placed on individual components, such as a single PMT cathode, by using the full component name (e.g., ``\texttt{Bottom\_PMT\_Window\_12}''). Alternatively, one can place the same radioactivity levels on multiple components using any part of the component name, e.g. the command ``\texttt{/LUXSim/source/set Bottom\_PMT\_Window AmBe 100 mBq/kg}'' will place the specified AmBe activity on all bottom PMT windows. The LUXSimManager keeps a record of the total activity for each component, and each component keeps the record of the sources it contains. In keeping with the LUXSim guiding principles, this functionality is automatically available for all components cast as a LUXSimDetectorComponent, requiring no additional coding on the part of a developer.

When an event is to be generated, Geant4 calls the LUXSimPrimaryGeneratorAction class, which in turn passes the call on to the LUXSimManager, along with the required pointers to the manager class. The LUXSimManager chooses one detector component to generate an event position; the component selection is random, but weighted by the total activity of each component. The manager passes the event generation pointers to the selected component.  The component selects a position within its own geometry to generate the decay. The component class also randomly chooses a source type, again weighted by the activities of all sources in that component. The component calls the appropriate LUXSimSource object, which specifies particle type, energy, and direction. Finally, with the particle position, type, energy, and direction specified, the source object calls the Geant4 generator, passing all the required primary particle information.

\subsubsection{Single-nuclide decays}

All single-isotope decays are handled in a single method of the LUXSimSource class. This method takes advantage of the built-in Geant4 Radioactive Decay Manager (G4RDM), which contains all the isotope decays from the Evaluated Nuclear Structure Data Files~\cite{Tuli1987}. The G4RDM provides $\alpha$, $\beta$, and $\gamma$ decays from both excited- and ground-state nuclei. The G4RDM also takes as input the range of mass number and charge over which a nucleus is allowed to decay, allowing for a decay of just the parent nucleus, decays all the way to a stable nucleus, or somewhere inbetween.

The macro command used to specify the decay must include the atomic mass and number. For example, to create \iso{40}{K} decays, one would use ``SingleDecay\_40\_19''. As with all macros, this command is parsed by the LUXSimManager. The generator uses G4GPS to define the particle as the requested isotope in an uncharged, ground state. When using this macro command, LUXSim sets the nuclear decay limits to just the parent isotope, rather than any possible extended chain of decays. After giving the isotope zero energy, the event is started with the usual call to the Geant4 method GeneratePrimaryVertex.

\subsubsection{\iso{232}{Th} and \iso{238}{U} decay chains}
\label{sss:UThDecayChains}

A common approach to decay chain generation within Geant4 is to simply use the G4RDM. One particular concern in decay chains is position correlation---if the radionuclides are trapped within a bulk material, then a daughter decay should occur at the same location as the parent decay. If full position correlations are required, the chain is allowed to decay to the last, stable isotope. If position correlations are not required, single decays may be created at random throughout the decay chain. Individual experiments may slightly alter these basic approaches based on detector response, activity levels, and so forth.

These two approaches of correlated-position and uncorrelated-position decays each have disadvantages that were avoided within LUXSim. In the correlated-position approach, a required input for accurate decay chain generation is the age of the source. If the source age is large compared to the time between decays, many unnecessary decays are simulated before the age of the source is reached. This requires computation time on simulated events that will ultimately be discarded from the analysis. Another disadvantage is that multiple traversals of the decay chain result in temporally interspersed events. For example, the \iso{224}{Ra} decay from a traversal of the \iso{232}{Th} decay chain may occur before the \iso{228}{Ac} decay from another chain. This becomes an issue if two decays from different traversals of a decay chain occur within the event pileup window of the detector. Thus the events recorded to disk must be post-processed to arrange them in chronological order, and the computation overhead to time-order millions to billions of events may be substantial.

The uncorrelated-position approach, using random decays within the chain, avoids both the problem of simulating unnecessary decays and having to time-order the events in post-processing, but it unfortunately loses the position correlations between individual decays. Extrapolations can be performed to minimize the effects of losing position correlations. These extrapolations, however, require detailed knowledge of the detector response and specific source activity levels. The resulting decay chain generator is therefore single-purpose, and cannot be used in simulations of other detectors that do not have near-identical operation and response.

\begin{figure}[t]
\centering
\includegraphics[width=8.5cm] {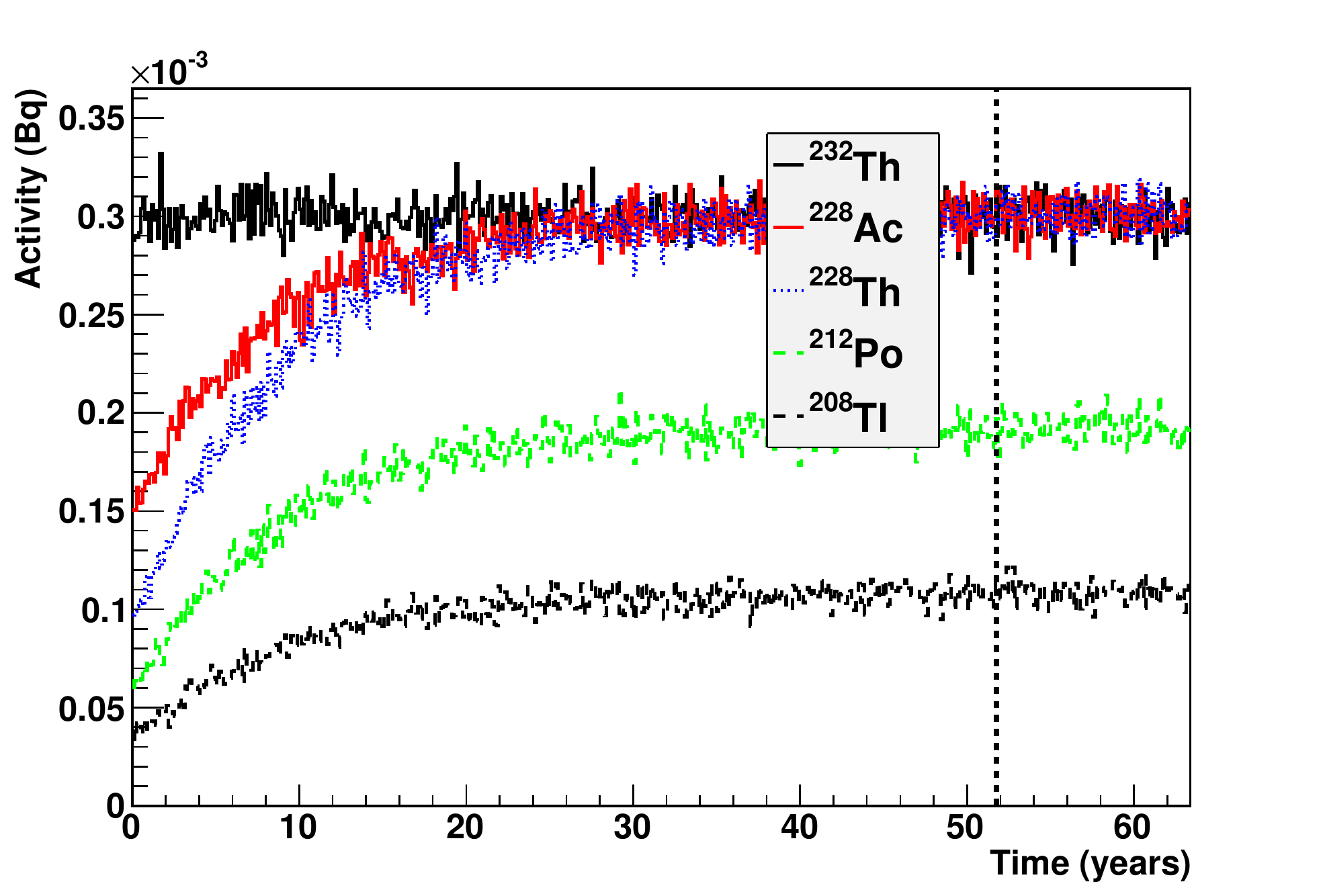}
\parbox{8.5cm}{\vspace{5pt}\caption{\small{Decay activities of selected isotopes within the \iso{232}{Th} decay chain. In this figure, the source has an age of 1 half life of \iso{228}{Ra}, or 5.75~years. The vertical dashed line marks 10 half lives from a source age of 0 years. The source age can be arbitrarily set without additional computational requirements. Figure taken from Ref.~\cite{Kazkaz2011}.}}
\label{fig:DecayChain}}
\end{figure}

LUXSim avoids all these disadvantages by using a general-purpose decay chain generator that exhibits accurate timing and maintains position correlations, without simulating discarded events or requiring chronological post-processing. The LUXSim decay chain generator uses the source age and activity level of the parent isotope as inputs. The approach used combines analytic and stochastic methods to create a time-ordered record of all decays before a single event has been generated. This approach is fully documented in Ref.~\cite{Kazkaz2011}, including benchmarks and memory usage analysis. A sample plot of isotopic activity levels from the \iso{238}{U} chain is shown in Fig.~\ref{fig:DecayChain}.

\subsubsection{AmBe neutrons}

The AmBe generator in LUXSim produces neutrons with a representative spectrum along with the accompanying high-energy gamma rays. The 60-keV gamma rays associated with the decay of \iso{241}{Am} are not included in this generator. If such decays are required, a basic americium source can be loaded into a detector component, along with this nominal AmBe source, with the appropriate activity ratios. Self-shielding effects can be created naturally by creating a detector part made of americium-beryllium material, and loading that volume with the AmBe source. If a point source is desired, the volume can be made nonphysically small (e.g., an angstrom in extent).

\begin{figure}[t]
\centering
\includegraphics[width=8.5cm] {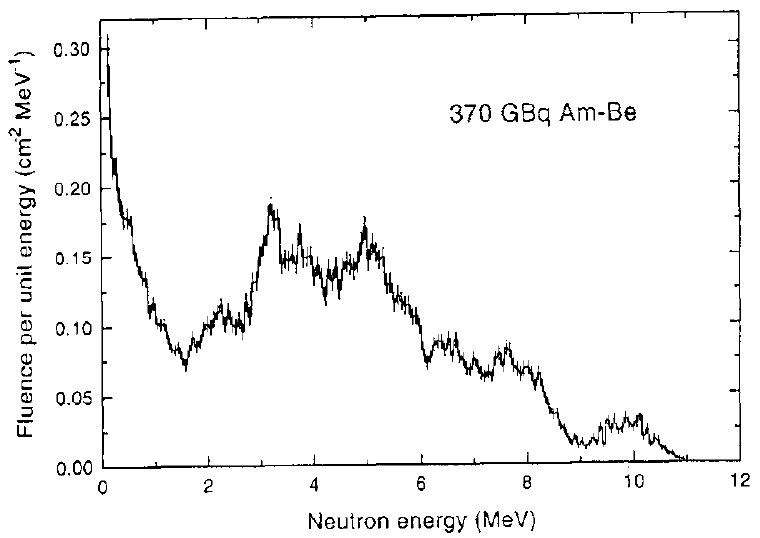}
\parbox{8.5cm}{\vspace{5pt}\caption{\small{Spectrum of neutrons from an AmBe source. This spectrum was digitized and used in the LUXSim AmBe event generator. Figure taken from Ref.~\cite{Marsh1995}.}}
\label{fig:AmBeSpectrum}}
\end{figure}

The neutron spectrum comes from Marsh {\it et al.}~\cite{Marsh1995}, and is reproduced in Fig.~\ref{fig:AmBeSpectrum}. This spectrum was digitized and normalized in order to create a cumulative distribution function (CDF) with a neutron endpoint energy of 11~MeV. Within the generator, a random number between 0 and 1 is generated, and the CDF is used to associate that random number with an energy. A linear interpolation is used between the discrete energy values. This method of spectrum sampling was used because the sample space is somewhat sparsely populated, and requires creating just one random number that is converted to a neutron energy.

The neutron spectrum can depend greatly on the source geometry, especially for the lower-energy neutrons caused by break-up reactions. Even so, this neutron spectrum was selected as representative of the large majority of AmBe sources available, and any deviation from this reference source would have to be measured for each individual source.

($\alpha$, n) reactions within the AmBe source can be accompanied by 0, 1, or 2 gamma rays, depending on the energy level of the resulting \iso{12}{C} nucleus. The associated gamma energies were taken from the NuDat data base~\cite{Ajzenberg1990}, and the \iso{12}{C} energy level structure from the Isotope Explorer~\cite{Isotope2010}. The number of gammas produced in coincidence with a neutron is a function of the neutron energy, and comes from Geiger and Zwan~\cite{Geiger1975}. Neutrons with energy up to 0.5~MeV are not accompanied by gamma rays, as they are classified as break-up neutrons. Neutrons with energy between 0.5 and 1.9~MeV are associated with a \iso{12}{C} nucleus in the 7654-keV energy level, and this nucleus relaxes via emission of 3215- and 4439-keV gamma rays. If the neutron has energy between 1.9 and 6.0 MeV, the \iso{12}{C} nucleus is in the 4439-keV state, and decays via emission of a single gamma ray. Neutrons above 6.0~MeV are not accompanied by gamma rays.

When two gamma rays are emitted, the LUXSim AmBe generator takes into account the angular correlations between these gamma rays. The 7654-keV level is a $0^+$ state, the 4439-keV is $2^+$, and the ground state is again $0^+$, meaning both nuclear relaxations are electric quadrupole transitions. Because the parent \iso{12}{C} nuclei are considered to be non-polarized, a random direction is selected, and the individual gamma emission angles distributed from that random angle via the appropriate Legendre polynomial equation.

\subsubsection{Fission generator}

The fission generator is modeled on a \iso{252}{Cf} source. The fission generator does not reproduce a full \iso{252}{Cf} source, but only creates neutrons and gammas associated with spontaneous \iso{252}{Cf} fission. Similar to the AmBe generator, a full \iso{252}{Cf} source can be created by loading a simple \iso{252}{Cf} source onto a source volume along with this fission generator in the appropriate ratio. Currently, only the \iso{252}{Cf} generator is available in LUXSim. Others can be added using the approach and references contained in this section.

\begin{figure}[t]
\centering
\includegraphics[width=8.5cm]{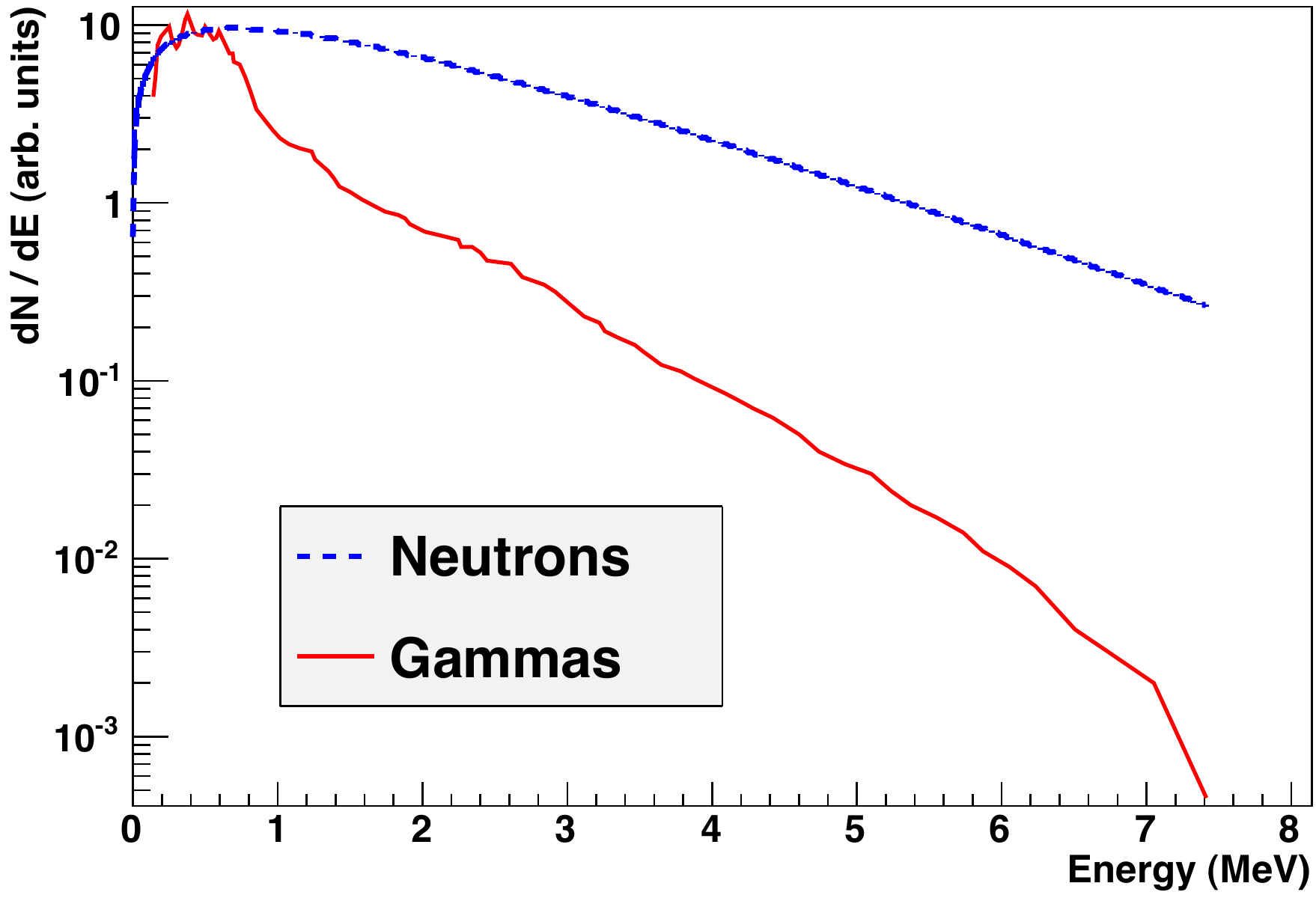}
\parbox{5in}{\vspace{5pt}\caption{\small{Neutron and gamma ray energy spectra from spontaneous fission decays of \iso{252}{Cf}. The neutron spectrum comes from Eq.~\eqref{eq:FissionSpectrum}, and the gamma spectrum is from Ref.~\cite{Verbinski1973}. The curves are not correlated to each other in amplitude, and merely display their relative shapes.}}
\label{fig:CfFission}}
\end{figure}

The multiplicity of the neutrons has a mean value of 3.757, as reported in Valentine~\cite{Valentine1999}, while the energy of the individual neutrons is described by a Watt spectrum:

\begin{equation}
\label{eq:FissionSpectrum}
dN/dE = e^{\left ( -E / 1.209 \right )} \mbox{sinh}\sqrt{0.836 E}
\end{equation}

\noindent
where $E$ is in units of MeV. The neutron spectrum is shown in Fig.~\ref{fig:CfFission}. The parameters for Eq.~\eqref{eq:FissionSpectrum} come from Mannhart~\cite{Mannhart}. The energy range extends from 1~meV to 15~MeV. Because the spectrum takes on an analytic form, selecting from it at random is handled differently than the neutron spectrum of the AmBe generator. The maximum value of Eq.~\eqref{eq:FissionSpectrum} is slightly below 0.48, so a random coordinate is chosen with x ranging from 0 to 15 and y ranging from 0 to 0.48. If this coordinate point lies below the neutron energy curve, that energy is selected as the event's neutron energy. Otherwise, a new coordinate point in the same (x, y) range is chosen at random. This method of spectrum sampling was used because the available parameter space is roughly 50\% filled. If the parameter space were appreciably less than 50\% filled, a CDF would have been used instead, in the manner of sampling from the AmBe neutron spectrum. 

The neutron multiplicity of a fission event defines the total energy of the gammas released in the fission event via the equation

\begin{equation}
\label{eq:FissionGammaEnergy}
E_{Tot,\gamma} = \left ( 2.51 - 1.13 \times 10^{-5} Z^2 \sqrt{A} \right ) \nu + 4.0 \mbox{~MeV}
\end{equation}

\noindent
where $Z$ and $A$ are the atomic number and weight of \iso{252}{Cf}, and $\nu$ is the number of emitted neutrons. Eq.~\eqref{eq:FissionGammaEnergy} is an empirical fit to the data, and there is no physical justification for its form~\cite{Valentine1999}. Valentine collates a few measurements of the average total gamma energy released in the fissioning of \iso{252}{Cf}, and determines a best-fit value of $6.95 \pm 0.3$ MeV ($\sigma = 4.32\%$). We therefore apply a 4.32\% Gaussian spread to the total energy in any given simulated \iso{252}{Cf} event within LUXSim.

The average energy of the emitted gammas is calculated with the equation

\begin{equation}
\label{eq:AverageFissionGammaEnergy}
\left < E_\gamma \right > = -1.33 + 119.6~Z^{1/3} / A \mbox{~MeV}
\end{equation}

\noindent
To obtain the fission gamma multiplicity, the total energy is divided by the average energy for each event, and the resulting number is used as the mean of a Poissonian distribution, with an integer number of gammas determined stochastically from this distribution on an event-by-event basis.

The energy of the gammas is determined from the \iso{252}{Cf} spectrum available from Verbinski {\it et al.}~\cite{Verbinski1973}. The fission gamma spectrum extends to roughly 7~MeV, and within LUXSim, this spectrum is sampled at random for every gamma, with the restriction that the total energy must add up to that determined via Eq.~\eqref{eq:FissionGammaEnergy}. The gamma spectrum is shown in Fig.~\ref{fig:CfFission}. Because the parameter space for gamma energies is somewhat sparse, we used a CDF and a single random number to sample from the spectrum.

\subsubsection{Cosmic muons and spallation neutrons}
The cosmic muon and spallation neutron generator produces either muons or spallation neutrons spread randomly throughout the rock-cavern interface.  The equations governing particle characteristics discussed in this section come from Mei \& Hime~\cite{MeiHime}. The throughgoing muon flux is given as

\begin{equation}
I(h,\theta) = (I_1 e^{-h/\lambda_1} + I_2 e^{-h/\lambda_2}) \mathrm{sec}(\theta)
\label{eq:muon_theta}
\end{equation}

\noindent
where $I_1$ and $I_2$ are differential muon intensities at depth $h$, $\lambda_1$ and $\lambda_2$ are attenuation constants, $h$ is the slant depth ($h = h_0$ sec$(\theta)$, where $h_0$ is the vertical depth), and $\theta$ the slant angle. A flat over-burden has been assumed. For the Homestake site, $h = 4.3$ km water equivalent is used. Eq.~\eqref{eq:muon_theta} is a fit function to the available experimental data. The muon's incident angle is determined by sampling from equation Eq.~\eqref{eq:muon_theta}.  Once the angle has been assigned, the muon energy is sampled from

\begin{equation}
\frac{dN}{dE_\mu} = A e^{-bh(\gamma_\mu-1)} (E_\mu + \epsilon_\mu(1 - e^{-bh}))^{-\gamma_\mu}
\label{eq:muon_energy}
\end{equation}

\noindent
where A is a normalization constant, $E_\mu$ is the muon energy, and b, $\gamma_\mu$, and $\epsilon_\mu$ are parameters detailing energy loss through rock. Once the muons are generated, they are handled by the internal Geant4 code. The Geant4 code itself handles various interactions, including spallation neutron production.

Through the use of a macro command, users have the option of generating spallation neutrons as primary particles instead of cosmic muons. An event site is chosen at random throughout the cavern volume, and the neutron energy, angular distribution, and multiplicity are sampled from Mei \& Hime Eqs. (14) and (15), (16)-(18), and (19)-(22) respectively.  Although spallation neutrons are created over a lateral range extending meters away from the muon track, a single event site is appropriate for fast, order-of-magnitude calculations.

\subsubsection{Primary scintillation and ionization}
\label{sss:PrimaryScintIon}

The model of scintillation and ionization must be as accurate as possible to provide realistic calculations of the detector response. The basic installation of Geant4 incorporates a model for scintillation light production~\cite{PhysRefManual2010,ApplicationDevelopers2010}, although it has some deficiencies that were addressed within LUXSim. As discussed in the section, however, the ionization model is wholly inadequate for dual-phase detectors.

In standard Geant4, the number of scintillation photons created per unit of energy loss is set by the user for each material and particle type. Unfortunately, this ability to distinguish between particles only applies to protons, electrons, deuterons, tritons, alphas and a generic ion particle definition~\cite{ApplicationDevelopers2010,Gumplinger2011}. More specifically, xenon nuclei scintillation parameters cannot be uniquely defined in standard Geant4. In dual-phase detectors, the scintillation yield is also a function of the applied electric field; a strong electric field reduces recombination, but recombination of ionized electrons leads to scintillation. Geant4 does not allow for this E-field dependence. Furthermore, while the Geant4 scintillation model allows us to set the ratio of long and short decay constants as a function of particle type, it does not allow for variability with respect to particle energy. 

The Geant4 scintillation model depends entirely on user definitions. The parameters such as yields, decay constants, and output spectrum are not distributed with Geant4 itself. It falls to the user to evaluate all available publications to determine the best parameters for xenon scintillation. These parameters are entered in the form of arrays, with interpolations performed between data points.

In addition to scintillation deficiencies, the standard Geant4 physics models are incapable of tracking ionization electrons below 250~eV. Rather, if any number of low-energy electrons are created below this cutoff value, they deposit all their energy in a single point without any tracks being created. Because the ionization electrons created by a xenon recoil have energy in the 10-eV range~\cite{Gerber1972}, the electromagnetic physics models in Geant4 are not directly applicable to a dual-phase detector. Likewise, Geant4 does not allow for drifting electrons, as their energy is comparable to that of the initial ionization electrons~\cite{Biagi1999}.

To obtain a more accurate scintillation and ionization yield for a wide variety of situations in dual-phase xenon, the Noble Element Simulation Technique (NEST) code was developed~\cite{Szydagis2011} and implemented in LUXSim. NEST uses energy-, particle-, and field-dependent models to determine scintillation and thermal ionization yield. It is applicable to both electron and nuclear recoils with energies $O$(1~keV) to $O$(1~MeV), as well as varying electric fields from 0 to $\sim$10 kV / cm. NEST offers LUXSim and similar applications a recombination model that is vetted against all known available experimental xenon data, making additional user input unnecessary. The best comprehensive results were used in the scintillation and ionization yield equations, which replace the Geant4 approach of using arrays to define the scintillation parameters, and thus avoiding interpolation.

Through the use of NEST, LUXSim is able to create both light and charge yield in a realistic fashion under a wide variety of conditions and parameters. This will allow LUXSim to perform simulations of the full chain of events, from initial particle interactions in the liquid, to drifting electrons, to the production of secondary scintillation light.

%%%%%
\subsection{The physics list}
\label{ss:PhysicsList}

LUX is a low-background experiment, and like most such experiments, it will be located in an underground laboratory. The theoretical scale of nuclear recoil energies from WIMP interactions is in the 1-100~keV range, and as such, LUXSim employs physics models that extend to these lower energies. At the same time, however, underground experiments must also contend with high-energy cosmic muons that survive the kilometer-water-equivalent-scale overburdens. These tend to have energies in the hundreds of GeV range. LUXSim must therefore also be able to handle high-energy interactions. Neutrons creating nuclear recoils are a background to the WIMP signal, and they must be handled over a range of energies from thermal to hundreds of MeV. LUXSim must also be able to properly generate and handle optical photons, as well as radioactive decays.

LUXSim calls up the appropriate physics lists via the modern list factories available starting in GEANT version 4.9.2. For low-energy electromagnetic interactions, LUXSim makes use of the Livermore physics list. Details of the Livermore list, as well as other lists described in this section, are, unless stated otherwise, available in the Physics Reference Manual~\cite{PhysRefManual2010}, with implementation options described in the Users's Guide for Application Developers~\cite{ApplicationDevelopers2010}.

The high-energy models must handle interactions such as spallation events, elastic collisions, and short-lived particles and their decays. LUXSim makes use of the QGSP\_BIC\_HP list. These terms stand for, in order, ``Quark-Gluon String Precompound'', ``Binary Cascase'', and ``High-Precision''. QGSP provides string models for hadrons above 5-25~GeV, and parameterized models for lower energies. The Binary Cascade models are used for protons and neutrons below 10~GeV. The High-Precision models are data-driven models for neutron transport from 20~MeV down to thermalization. The QGSP\_BIC\_HP list does not handle relaxation of nuclei resulting from neutron captures. Processing of excited-state nuclei is handled via separate hadronic models.

The optical physics list was created in consultation with Peter Gumplinger, and is based on the {\it field04} extended example included as part of the Geant4 distribution. The optical physics list includes absorption, Rayleigh scattering, and boundary processes. It also allows for the generation of optical photons via either Cherenkov production or scintillation. Because a large number of optical photons can be generated with even modest energy depositions, the optical photon generation can be turned on or off at run time via LUXSim macro commands to speed up the simulation if optical physics is unnecessary. A simplified optical physics model is also available for selection at run time. The details of this model are covered in Section~\ref{ss:OpticalProperties}.

An often-used parameter within the Geant4 physics list is known as the ``cut value''. This is the energy below which secondary particles are no longer created. Because particles will have different ranges for the same energy depending on the medium, the cut value is set as a length rather than an energy, and can be set separately for different particles. The default setting within LUXSim is 5~$\mu$m for gammas, electrons, and positrons, and 100~$\mu$m for protons, $\alpha$'s, and generic ions. Using a short cut value can greatly increase the simulation run time. If tracking at the 5-$\mu$m level is not required, the cut value for gammas, electrons, and positrons can be set to 100~$\mu$m at run time via LUXSim macro commands. There is no apparent increase in simulation speed for increasing the cut value above 100~$\mu$m.

%%%%%
\subsection{Event processing}
\label{ss:EventProcessing}

Event processing is handled via the usual Geant4 methods UserSteppingAction, BeginOfEventAction/EndOfEventAction, and BeginOfRunAction/EndOfRunAction. Within the stepping action, individual steps, interactions, and energy depositions are stored in memory for later processing. While there can be a very large number of steps within a single event, the number rarely goes above hundreds of thousands, even for high-energy events. While this requires hundreds of megabytes of computer memory, it is easily within the capabilities of any modern desktop or laptop. The stepping action also kills particles and tracks under certain circumstances, if requested by the user (see Section~\ref{ss:DataRecording}).

The event action prints out periodic progress reports, and after all steps have been stored, it calls on the manager to determine what data from the full collection of any given event needs to be written to disk. Finally, the store of data is cleared from memory in anticipation of the next event. A great deal of the work of the run action has been incorporated into the manager methods, so the run action's only function is to print start and stop times to the screen.

%%%%%
\subsection{Data recording}
\label{ss:DataRecording}

LUXSim includes a specific class, LUXSimOutput, to process the step information that is recorded during an event. The output file itself is a custom-format binary file. It includes two groups of information. One group is the header which records the information about the run, and includes the following: 

\begin{itemize*}\vspace{-0.5\baselineskip}
\item A time/date stamp of when the simulation was run
\item The versions of Geant4 and LUXSim
\item The computer information, including name and operating system, on which the simulation was run
\item The macro command used in setting up the simulation
\item All the differences between that version of LUXSim and any changes that might have been made to the code directly
\item A detector component ID lookup table
\end{itemize*}

\noindent
The final item in this list, the ID lookup table, is included to save disk space. Volumes from which steps are recorded are not referred to by their string names, but by a numerical ID which is defined in the header.

\begin{figure}[t]
\centering
\includegraphics[width=15cm]{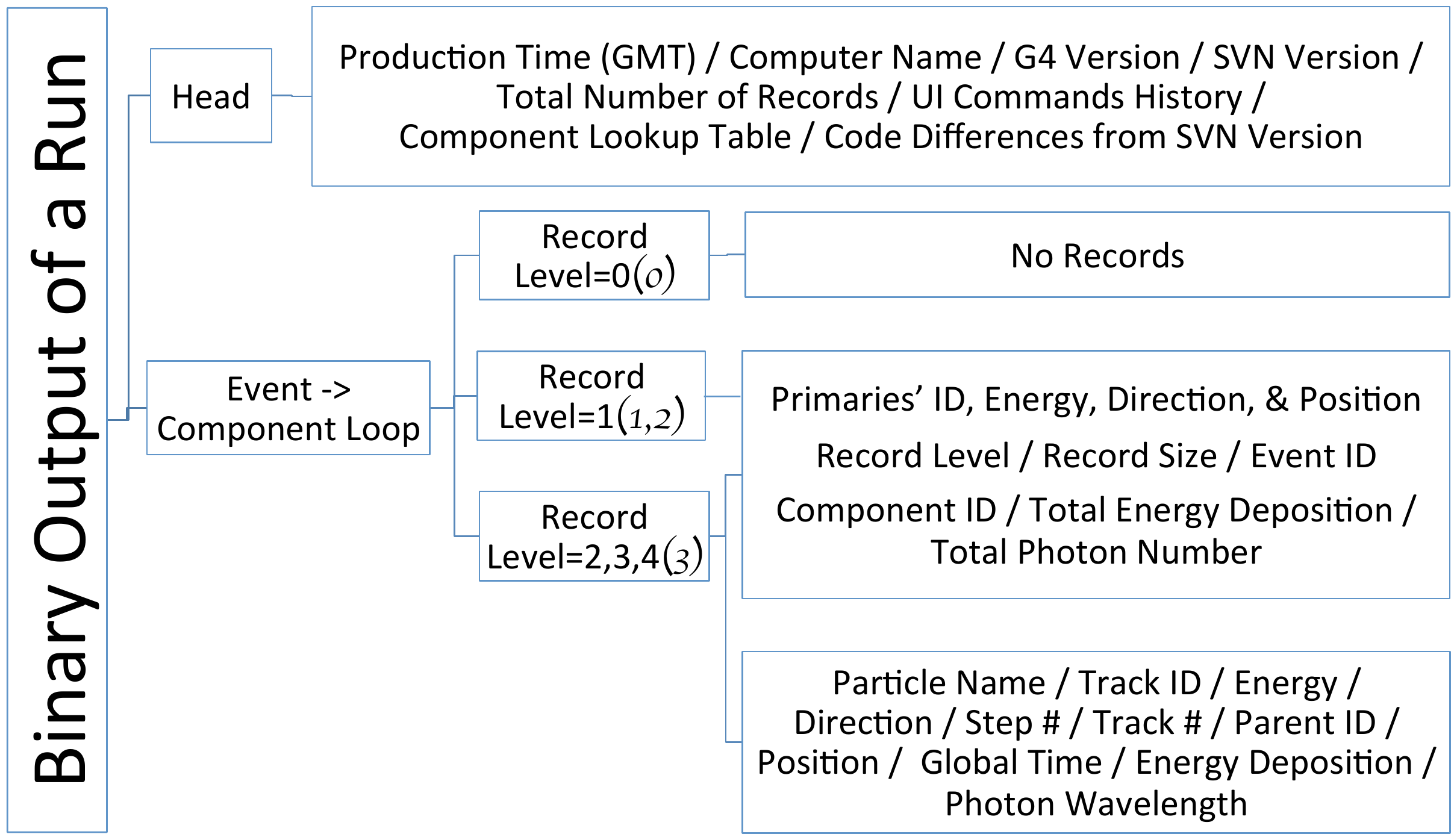}
\parbox{5in}{\vspace{5pt}\caption{\small{Flow chart of data recording in LUXSimOutput class.
The numbers in parentheses represent the optical record level. }}
\label{fig:DataFormat}}
\end{figure}

The vast majority of recorded information is from the results of the simulation itself, on a volume-by-volume basis, and is written after every event. LUXSimOutput determines how much information to record according to the record levels defined by the user in the macro command file. A flow chart for the data recording is shown in Fig. \ref{fig:DataFormat}. There are two independent record level categories, one for optical photons, and one for all other particles. This separation is necessary because optical photons are handled in ways distinct from other particles within Geant4, in terms of energy conservation, ionization, and fundamental physical processes. These record levels tell the output class what information to record for each detector component.

For Optical Photons:
\begin{itemize*}\vspace{-0.5\baselineskip}
\item Optical Record Level = 0  - Do not record (default)
\item Optical Record Level = 1  - Record just the total number of optical photons entering the volume, and kill the track so the photons do not propagate
\item Optical Record Level = 2  - Record just the total number of optical photons entering the volume but do not kill the tracks
\item Optical Record Level = 3  - Record all the information on the optical photons entering the volume and kill the tracks
\end{itemize*}

For normal particles:
\begin{itemize*}\vspace{-0.5\baselineskip}
\item Record Level = 0  - Do not record (default)
\item Record Level = 1  - Record just the total energy deposition in the current volume
\item Record Level = 2  - Record just the steps where energy was deposited
\item Record Level = 3  - Record all the steps, even those with no energy deposition
\item Record Level = 4  - Record all the information about the particle, then kill the track
\end{itemize*}

\noindent
Using record level 4 for ordinary particles, or optical record level 3 for optical photons, is used primarily for debugging purposes, or when we are not necessarily concerned about what happens inside a detector component, and simply want to know the flux of particles into the component.

%
%	Data Processing
%
\section{Post-Simulation Data processing}
\label{s:DataProcessing}

The geometries in LUXSim accurately recreate their physical counterparts, and the optical photon physics list provides accurate handling of the optical photons. To complete the simulation chain, we developed a detector response module to convert the LUXSim output files into detector-like data files, taking into account the response of the actual PMTs, electronic noise levels, pulse shaping, and triggering. This module is separate from the Geant4-based LUXSim code, and it is run on the output data from LUXSim. Our aim was to create data that is functionally identical to the actual experimental data, and can be processed with our standard experimental data analysis routines.

\begin{figure}[t]
\centering
\subfigure{\includegraphics[width=9cm]{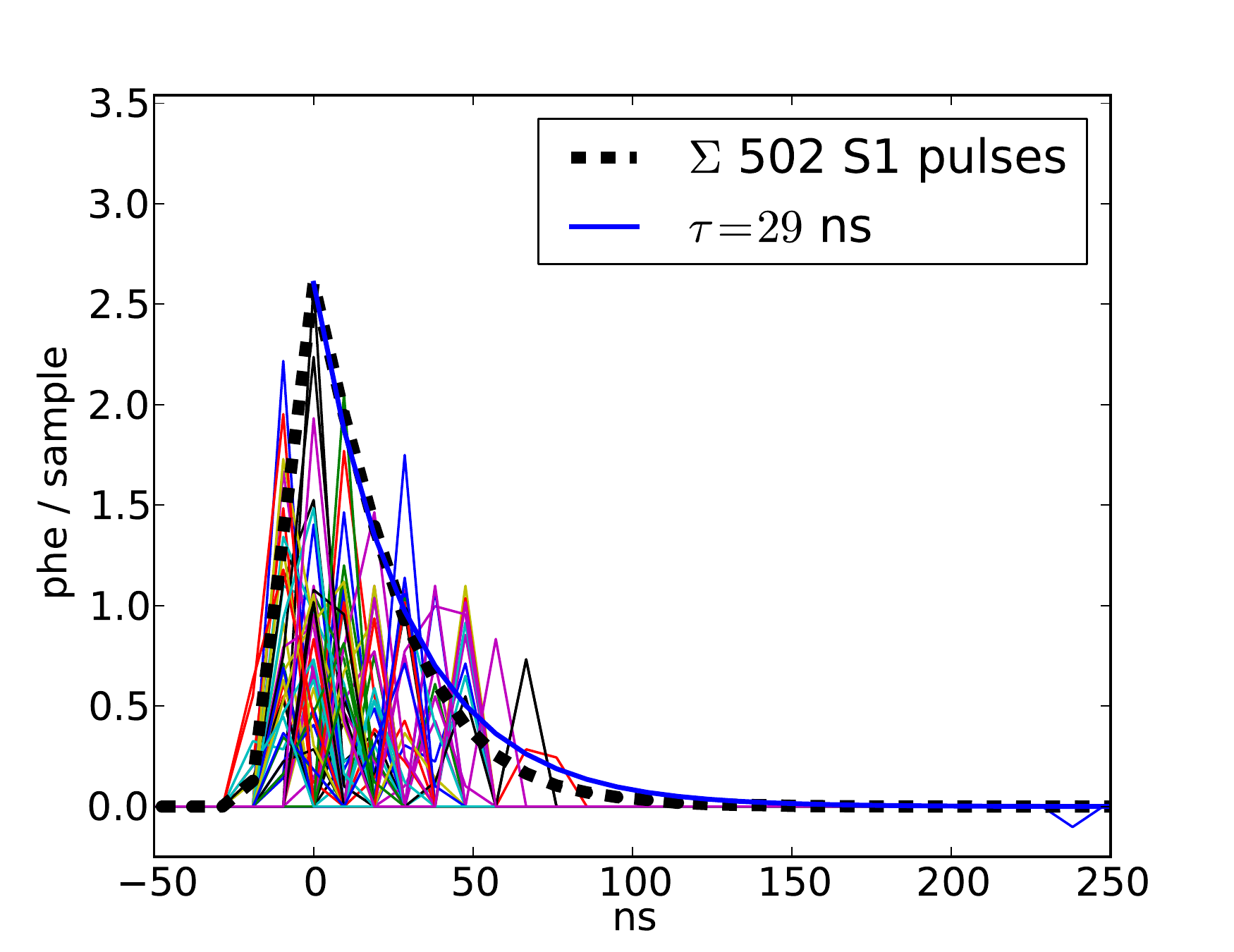}}
\subfigure{\includegraphics[width=8cm]{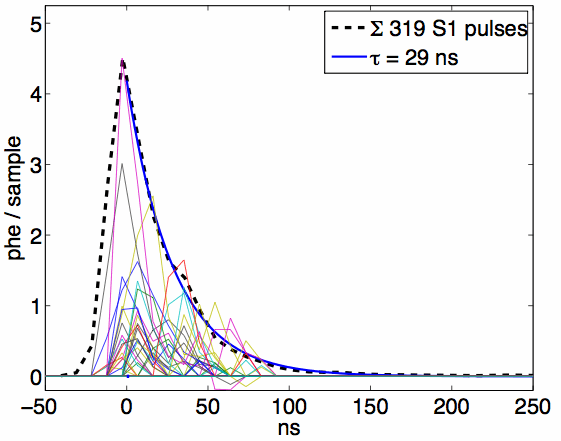}}
\parbox{5in}{\vspace{5pt}\caption{\small{The simulated LUXSim S1 response after post-processing (top), and a measured S1 signal from {\sc Xenon10} (bottom) \cite{Sorensen2008}. The horizontal axes show elapsed time since the trigger. The thick dashed line shows an average response for 502 (319) S1 pulses from LUXSim ({\sc Xenon10}) scaled by a factor of $2\times10^{-4}$ ($7\times10^{-4}$), with the response from the individual channels of single event in multi-colored lines. The scale factors are different because of differences in the energy deposition and the number of summed curves. A 29-ns decay curve is overlaid on both sum curves to show the measured scintillation relaxation time of liquid xenon.}}
\label{fig:SimulatedS1s}}
\end{figure}

The output of the simulated detector response has a binary format that mirrors the format of the detector data as written by the DAQ and is the starting point of the experimental data analysis. The data is stored as a collection of digitized waveforms, as shown in Fig.~\ref{fig:SimulatedS1s}, which can be used to develop, compare to, and test the collaboration's analysis tools. The digitized waveforms were compared to {\sc Xenon10} data because of the similarity of the detectors. Both LUX and {\sc Xenon10} are dual-phase xenon detectors with arrays of PMTs at the top and bottom, and thus the S1 signals from both detectors have similar characteristics. The simulated data stream allows us to test analysis and reconstruction algorithms.

Single-photon PMT responses are taken to be Gaussian in shape. A multiple-photon event is constructed by summing together single photoelectron responses for each photon that hits a PMT window. This results in the analog signal that is read from the full collection of PMTs. Simulation of amplifiers and shapers deliver the signal to a simulated digitizer. A frequency-independent noise of 155~$\mu$V RMS is added to the input of the digitizer. The simulated signals are based on the shaped single photoelectron response as measured from the LUX electronics. Figure~\ref{fig:SimulatedS1s} shows a simulated detector response to individual 30-keV electrons placed in the middle the liquid xenon. The LUXSim data was converted so that it mimics the experimental data that traversed the analog electronics and data acquisition system. The scintillation time constants are the cause of the tail of the S1 pulses.

%
%	Exercise
%
\clearpage
\section{Exercising the LUXSim software}
\label{s:Exercise}

With all the machinery of LUXSim in place, we exercised the software both to compare its results to experimental data, as well as to make predictions regarding the light collection of the LUX detector. These exercises are not intended to provide a full validation of the code, but rather to demonstrate basic functionality of the software, and to demonstrate that LUXSim provides reasonable results. LUXSim will be updated to incorporate experimental results as they become available.

This section is not intended as an exhaustive exploration of Geant4 performance, as such studies are available in the existing literature. Various groups have compared experimental data with Geant4's electromagnetics~\cite{Amako2005}, neutron spallation and transport code~\cite{Marino2008}, and hadronic shower models~\cite{Kiryunin2006}. Many other comparisons exist within the literature, and those referenced here are intended simply as examples from the larger body of work. The Geant4 collaboration itself maintains a list of publications covering model verification~\cite{GEANTVerification}. 

%%%%%
\subsection{LLNL single-phase detector}
\label{ss:LLNLSinglePhase}

\begin{figure}[t]
\centering
\includegraphics[width=10cm]{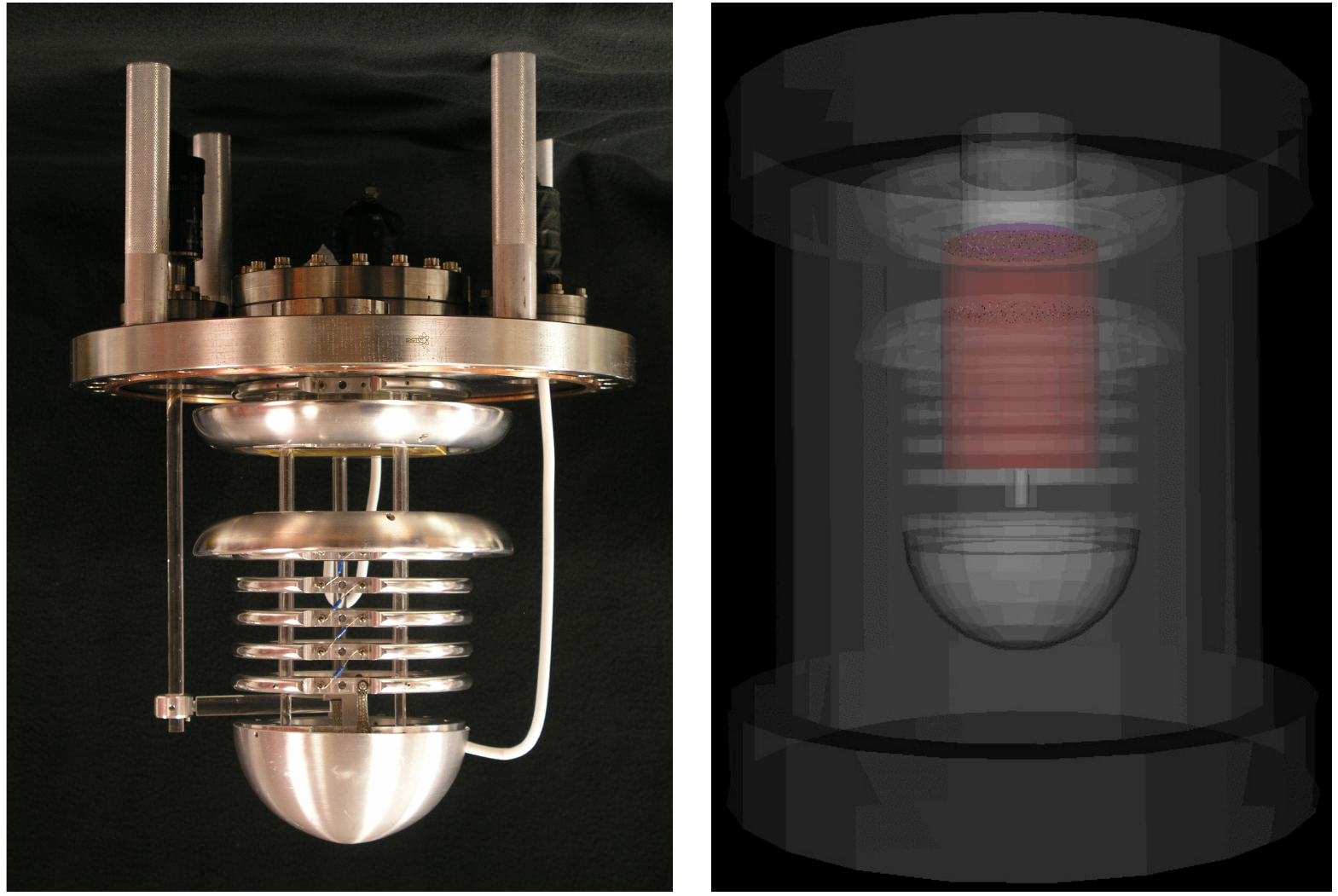}
\parbox{5in}{\vspace{5pt}\caption{\small{A photograph and a LUXSim rendering of the gas proportional scintillation counter used to study secondary scintillation light. The \iso{55}{Fe} source is located in a moveable collimator, shown at the end of the acrylic swing arm. The secondary scintillation volume is bounded by the large, toroidal field-shaping rings below the large steel flange. The acrylic support rods and swing arm have been left out of the simulation, as their effect on the collimated X-ray source is negligible. The collimator, however, is visible in both images.}}
\label{fig:LLNLSinglePhaseDetectors}}
\end{figure}

We have compared experimental and simulation data of an argon/nitrogen gas proportional scintillation counter with an \iso{55}{Fe} X-ray source. This detector was a single-phase system used as a testbed to study secondary scintillation, similar to the secondary signal produced by the LUX detector. The details of operation of the detector and data analysis methods are described in Ref.~\cite{Kazkaz2010}. Fig.~\ref{fig:LLNLSinglePhaseDetectors} shows a photograph of the experimental setup, and the corresponding LUXSim geometry.

The \iso{55}{Fe} source emits 6-keV X-rays which interact via the photoelectric effect, ejecting 3 keV electrons. The excited argon atoms relax via emission of either Auger electrons or by secondary X-ray emission. In the latter case, the X-rays may escape the detector, leaving behind a total energy deposition of only 3~keV. The S1 signals from these energy depositions were too weak to observe, but we constructed a spectrum from the S2 signal.

\begin{figure}[t]
\centering
\includegraphics[width=10cm]{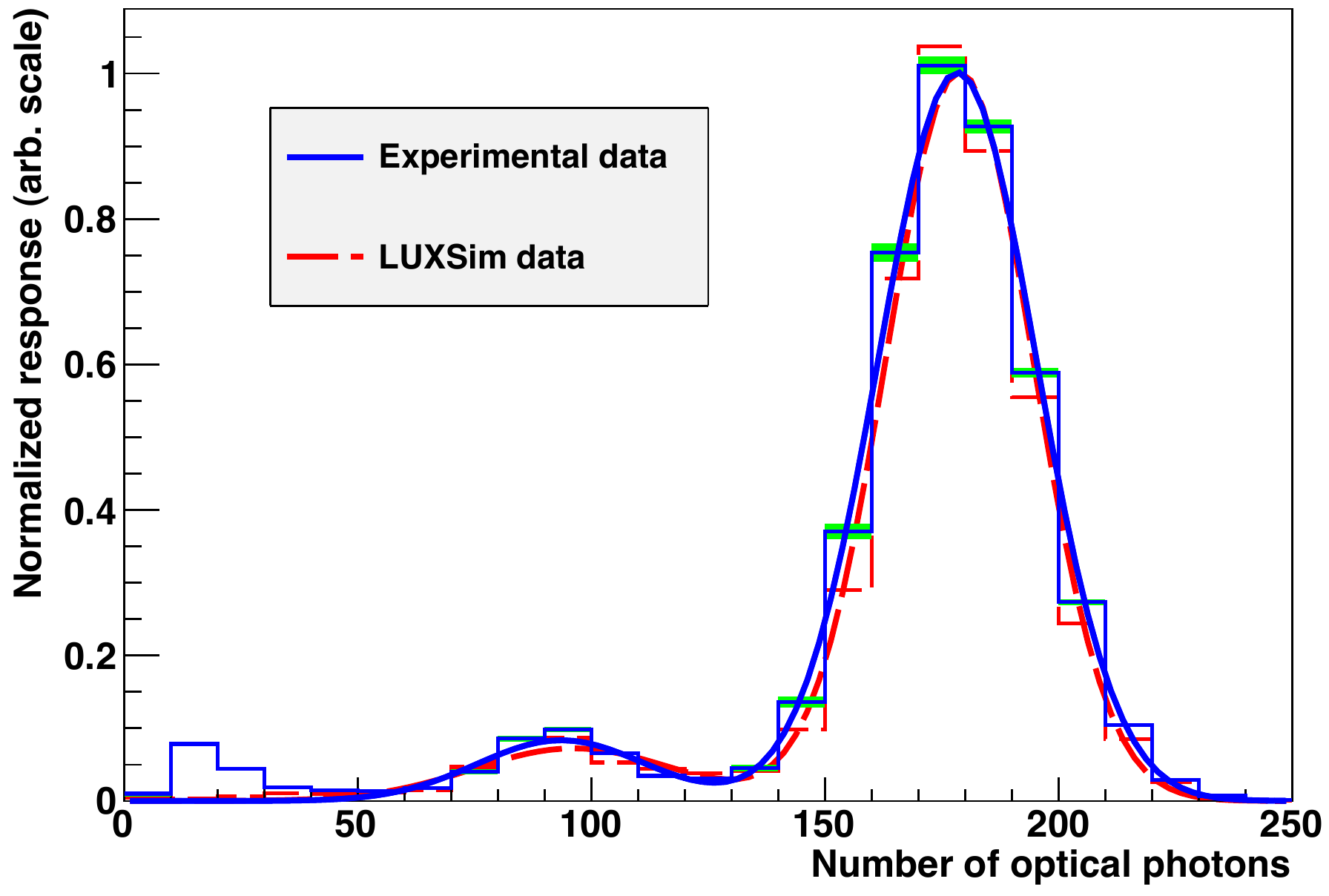}
\parbox{5in}{\vspace{5pt}\caption{\small{Comparison of the experimental and LUXSim S2 spectra from an \iso{55}{Fe} source. The two main features are a large peak near 180 optical photons corresponding to a $\sim$6-keV energy deposition, and a small peak near 90 optical photons corresponding to the $\sim$3-keV escape peak. The shaded regions show the uncertainty in the experimental data, and the smooth lines on each curve are fits to the data.}}
\label{fig:SinglePhaseEnergyComparison}}
\end{figure}

In the experimental data, we were able to observe a spectrum with two main features: a large peak corresponding to the 6~keV X-rays, and a smaller peak corresponding to escape events. These peaks were generated by LUXSim as well, in good agreement with experimental data (see Fig.~\ref{fig:SinglePhaseEnergyComparison}). To get the centroids of the experimental and simulation 6-keV peaks to match, the number of scintillation photons emitted / mm / ionization electron had to be set to 0.146. This was the only free parameter in the fit---the peak widths and relative amplitudes between the primary and escape peaks were predicted by LUXSim. The simulation did not include a quantum efficiency cut, and given the PMT's quantum efficiency of 27\% at the characteristic nitrogen range of 340-360~nm, we estimated the number of scintillation photons emitted in the physical detector to be approximately 0.54 / mm / ionization electron.

For this comparison between LUXSim and the experimental results, we constructed a rudimentary S2 signal generator by producing optical photons along a vertical line through the S2 volume. The $(x, y)$ position of this line was based on the location of energy depositions in the active gaseous volume, while the $z$ extent was defined by the top and bottom of the S2 volume. We were not able to make use of the NEST code described in Section~\ref{sss:PrimaryScintIon} because the medium in the single-phase detector was argon and nitrogen, rather than xenon. The resolution of the Monte Carlo curve was based solely on the number of optical photons detected. This implies that the width of the 6~keV peak in the experimental data is dominated by counting efficiency, in agreement with existing literature~\cite{Policarpo1970}. We conclude that maximizing light production and collection is the most effect means of improving the detector's resolution.

%%%%%
\subsection{Light collection in LUX}
\label{ss:LightCollection}

Because scintillation light is generated isotropically, most of the S1 light will at some point reflect off the walls, making detector response highly dependent on wall reflectivity. LUXSim was used to characterize scintillation photon reflectivity properties from the PTFE walls of the detector.

\begin{figure}[b]
\centering
\includegraphics[width=10cm]{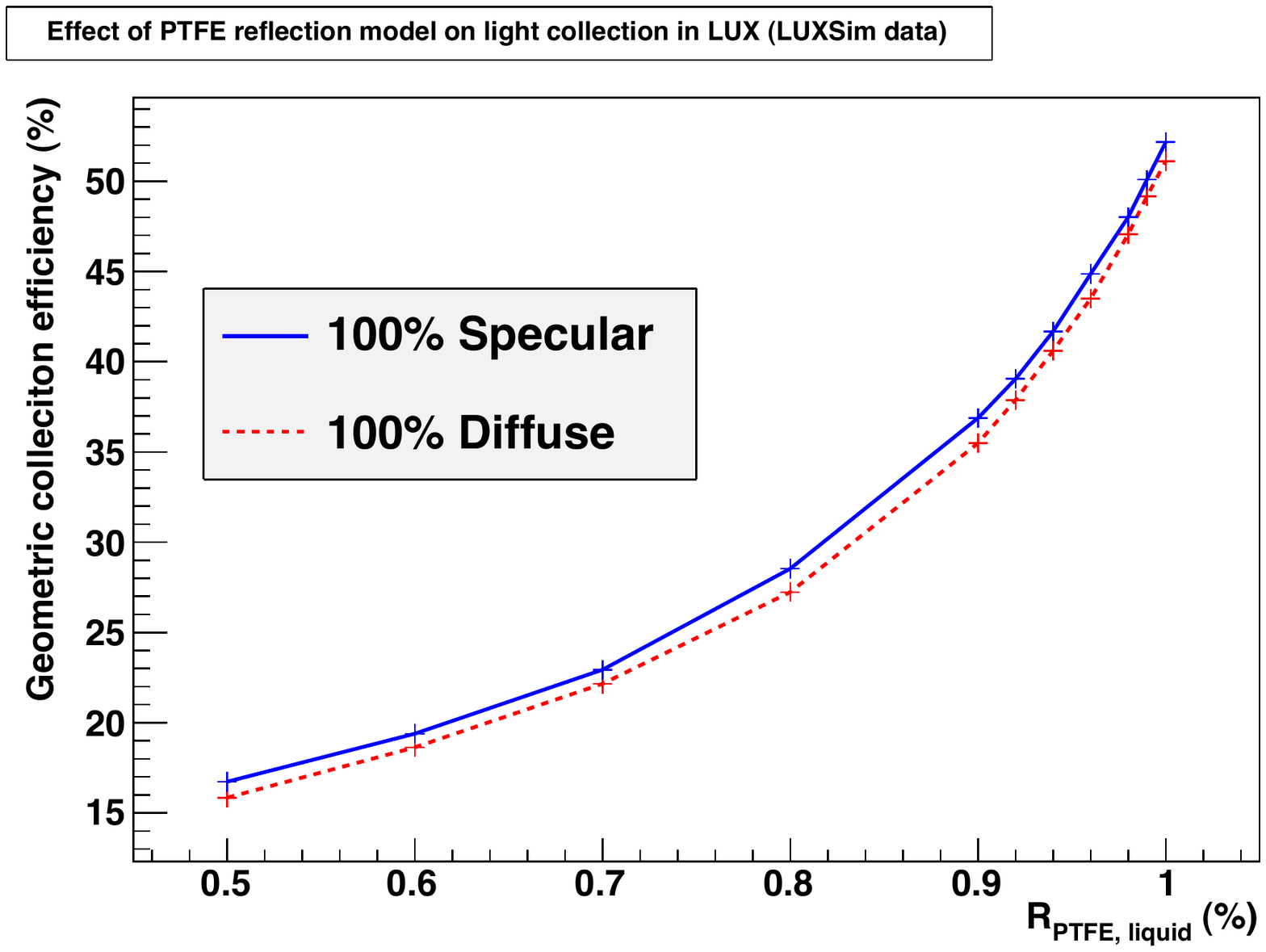}
\parbox{5in}{\vspace{5pt}\caption{\small{LUXSim calculation of the geometric light collection efficiency as a function of total PTFE reflectivity. The two curves represent extreme reflection models: 100\% diffuse and 100\% specular. The reflectivity of the PTFE plays a much greater role in overall efficiency than the reflectivity model used. The largest ratio between the curves, $\sim$1.05, is attributed to the effects of Rayleigh scattering in the liquid, which reduces the differences between specular and diffuse reflection. The uncertainty on the data points is less than 1\% (relative), and is therefore much smaller than the size of the data points.}}
\label{fig:CollectionEfficiency}}
\end{figure}

One question currently being studied is the effect on the total light collection of diffuse versus specular reflection from the PTFE walls. Silva {\it et al.} have developed a model governing precisely this issue based on measurements of reflectivity for PTFE for various manufacturing and processing methods~\cite{Silva2010a}. Though this model is available, we can use LUXSim to explore extreme cases of reflectivity, such as the effects of 100\% diffuse versus 100\% specular reflection on light collection. Figure~\ref{fig:CollectionEfficiency} shows the difference in geometric light collection efficiency for these two extreme cases. While LUXSim will strive to incorporate the best model of reflection available, the largest ratio between the two curves is about 1.05, demonstrating that the overall reflectivity of the PTFE has a much larger effect on light collection than the specific model of reflectivity used. In particular, the curve increases most strongly when reflection goes above 90\%. This strong response at high reflectivity is associated with individual optical photons bouncing multiple times from the PTFE walls. For example, consider an optical photon that bounces five times from the PTFE walls. With a reflectivity of 90\%, the chance of absorption is 41\%. This chance of absorption drops to 23\% for 95\% reflectivity, and just 5\% for 99\% reflectivity.

The curves in Figure~\ref{fig:CollectionEfficiency} were generated using the default optical parameters described in Section~\ref{ss:OpticalProperties}. The relatively small difference between the two curves in this figure are understandable from the standpoint of the 30-cm Rayleigh scattering length in the liquid xenon. The randomness introduced by diffuse reflection is comparable to the randomness introduced by Rayleigh scattering, such that even with a perfect, specular reflector, the optical photons tend to scatter multiple times along their path length.

A second study of light collection determined the qualitative effects of interaction location on detector response. If scintillation light is created close to a bank of photomultiplier tubes, the solid angle subtended by the tubes is relatively high, leading to a larger signal response than if the energy deposition were farther from the PMTs. This effect is counter-acted in part by the grid planes being most transparent at normal incidence, which implies that a higher percentage of light generated close to the PMTs would be absorbed by the individual grid wires. Complicating these issues are effects such as total internal reflection off the liquid / gas boundary, Fresnel reflection off the PMT windows, the aforementioned reflectivity models and scattering length, and the effects of discretized active PMTs.

\begin{figure}[htb]
\centering
\subfigure{\includegraphics[width=8.5cm] {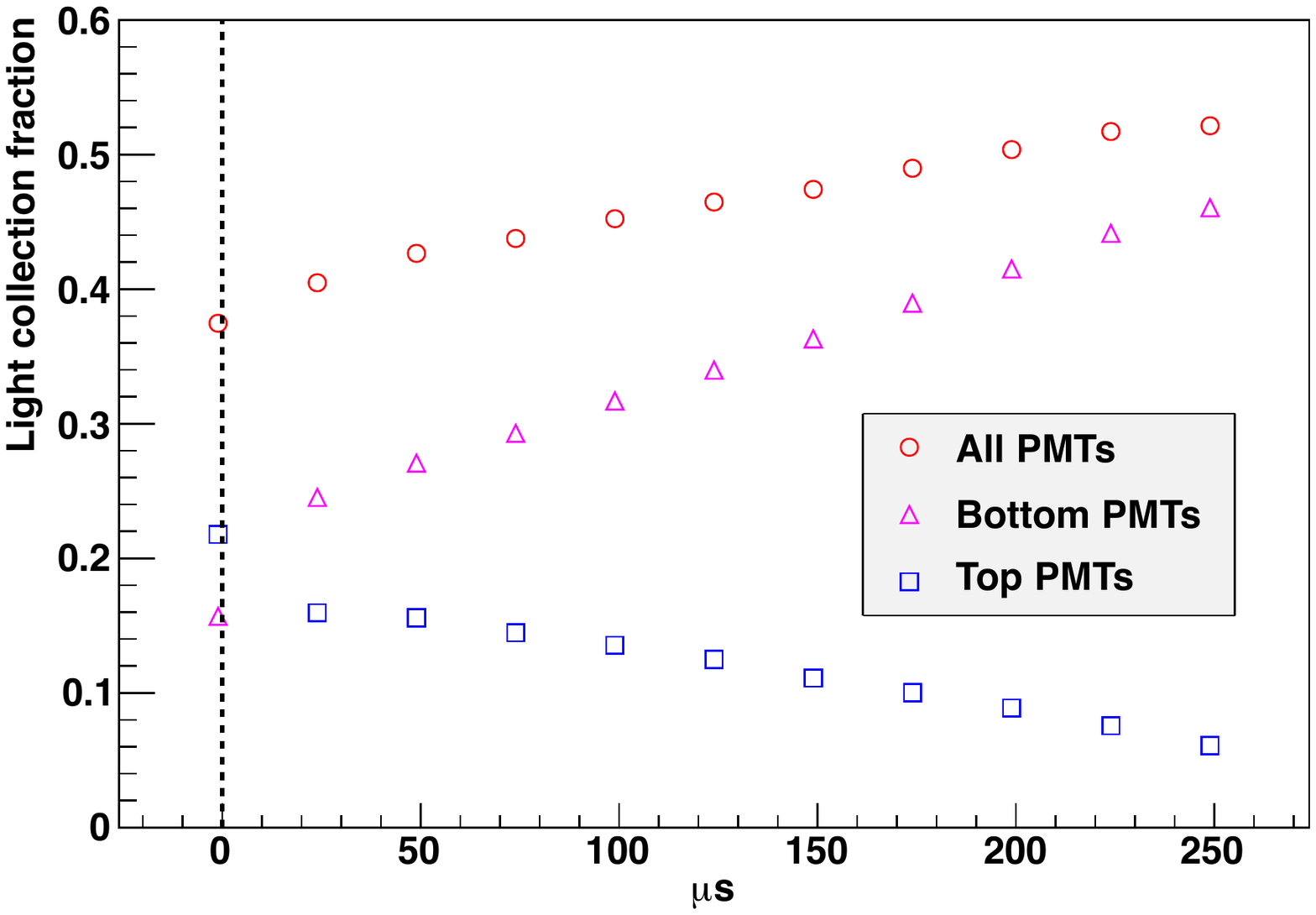}}
\subfigure{\includegraphics[width=8.5cm] {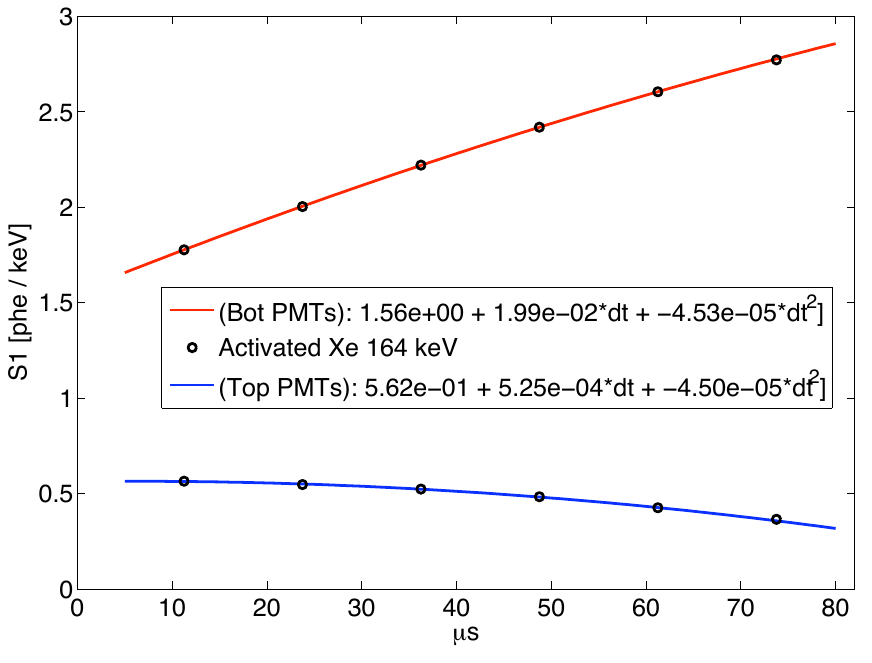}}
\parbox{8.5cm}{\vspace{5pt}\caption{\small{Relative light collection versus drift time for the {\sc Xenon10} detector (bottom)~\cite{Sorensen2008}, and as calculated for the LUX detector by LUXSim (top). The maximum drift time is greater in the LUX detector than the {\sc Xenon10} detector because of the greater height of the active liquid xenon volume. In the top plot, the dashed vertical line represents the liquid / gas interface, and the points that fall near this line were actually slightly above this interface, which explains the discontinuous trend in the data. Compare the triangles (squares) in the top plot to the upper (lower) curve in the bottom plot.}}
\label{fig:CollectionVsPosition}}
\end{figure}

Figure~\ref{fig:CollectionVsPosition} shows the difference in light collection as a function of drift time. To obtain the drift time, we assumed a typical drift speed of 2~mm~/~$\mu$s~\cite{Dahl2009}. The drift time is longest when the S1 scintillation was created at its lowest vertical position. Thus a drift time of 0~$\mu$s implies deposition in the gas volume, while a drift time of 250~$\mu$s implies scintillation created 50~cm below the liquid surface, near the bottom bank of PMTs. As we would expect, the closer to the bottom the energy deposition, the higher the light collection on the bottom PMTs, and the lower the collection on the top PMTs. As in Fig.~\ref{fig:SimulatedS1s}, the {\sc Xenon10} detector provides an apt comparison because of the detector similarities. Within LUX, these response curves will be measured {\it in situ} with appropriate calibration sources to generate a detector response map. We have implemented this level of detail in LUXSim to better understand the systematic effects of light collection that can be used to improve LUX or future detectors.

\clearpage

%
%	Expansion to other low-backround experiments
%
\section{Expansion to other low-background experiments}
\label{s:Expansion}

Because of its object-oriented nature, Geant4 provides great power for code re-use and compartmentalization. LUXSim capitalizes on this framework by respecting object-oriented programming practices. As a result, it is relatively simple to create a new geometry within LUXSim. Fully-developed Geant4 simulations may utilize custom methods and functions that control the geometry of the simulation. While these individual cases may require the custom code to be incorporated into LUXSim, the procedure to integrate an existing geometry is given by these steps:

\begin{enumerate*}\vspace{-0.5\baselineskip}
\item Perform a global search-and-replace in the existing geometry code base, replacing all instances of ``G4PVPlacement'' with ``LUXSimDetectorComponent''
\item Port over any macro commands specific to the existing geometry
\item Port over any custom event generators if they do not already exist within LUXSim
\item Alter the existing LUXSim geometry code base to reference the new geometry
\end{enumerate*}

\noindent
This last step is a very simple procedure, typically involving editing only 5-10 lines of code in total. With these four steps completed, users would be able to control the running of the simulation using the standard LUXSim macro commands. These commands automatically allow the user to set up component-centric event generation and recording at run time, without recompilations between simulations. The header information would be automatically written to the output file, greatly aiding reliability and reproducibility.

As additional experiments use LUXSim, we will incorporate additional physics models and materials, expanding the range of applicability. For example, it would be straightforward to incorporate optical parameters specific to argon and neon, making LUXSim appropriate for use in all noble-element WIMP searches. By simply porting in specific geometries and their associated material definitions, LUXSim can provide a very powerful simulations package for new or existing neutrinoless double-beta decay experiments.

The long-range plan for LUXSim includes making the code base publicly available, after the vetting process is complete. Until the code is available for download, inquiries for code access may be sent to the corresponding author.

%
%	Summary
%
\section{Summary}
\label{s:Summary}

LUXSim is an object-oriented simulations package based on Geant4. It involves expanding on the Geant4 classes to make Geant4 more immediately useful for low-background simulations. This expansion includes recording component-specific data, which allows for multiple source types and activities, various record levels, and automatic registration with the manager class. Newly-developed classes allow for a novel approach to event generators, allow for time-sequential processing, and minimize simulation time, file size, and post-processing.

We have presented an overall philosophy for guiding the development of the simulation software, clarifying and optimizing workloads, for both users and developers. These guiding principles were based primarily on ease of use, reproducibility, and physics requirements specific to low-background detectors with signals in the 1-keV to 10-MeV energy range. We have shown how the software architecture was guided by these principles, taking advantage of the object-oriented nature of C++ and Geant4 to make the code scalable while at the same time reducing the likelihood of coding errors.

We described the various subsystems and how they participate in the information flow within the simulation. We included details of event generation and recording, the physics models available within the simulation, and options for operating in a simplified physics mode for debugging purposes. We have also included geometry examples based on the LUX detector and associated projects.

%
%	Acknowledgements
%
\section{Acknowledgements}
\label{s:Acknowledgements}
We would like to thank Peter Gumplinger for assistance on developing an appropriate optical physics model. We would also like to thank the developers of Geant4 for creating such a powerful, scalable Monte Carlo software package.

This work was partially supported by the U.S. Department of Energy (DOE) under award numbers DE-FG02-08ER41549, DE-FG02-91ER40688, DOE, DE-FG02-95ER40917, DE-FG02-91ER40674, DE-FG02-11ER41738, DE-FG02-11ER41751, the U.S. National Science Foundation under award numbers PHYS-0750671, PHY-0801536, PHY-1004661, PHY-1102470, PHY-1003660, the Research Corporation grant RA0350, the Center for Ultra-low Background Experiments at DUSEL (CUBED), and the South Dakota School of Mines and Technology (SDSMT). We gratefully acknowledge the logistical and technical support and the access to laboratory infrastructure provided to us by the Sanford Underground Research Facility (SURF) and its personnel at Lead, South Dakota.

This work was performed under the auspices of the U.S. Department of Energy by Lawrence Livermore National Laboratory under Contract DE-AC52-07NA27344. Funded by Lab-wide LDRD. LLNL-JRNL-487572.

%
%	Bibliography
%
\bibliography{apssamp}% Produces the bibliography via BibTeX.

\end{document}